%% file: main.tex
\documentclass[conference,compsoc,letterpaper]{IEEEtran}
\IEEEoverridecommandlockouts

\pdfoutput=1

\usepackage{shortcuts}
\usepackage{booktabs}
\usepackage{fontawesome}

\renewcommand{\IEEEauthorrefmark}[1]{\textsuperscript{#1}}

\begin{document}

\title{Robust Backdoor Detection for Deep Learning via \\Topological Evolution Dynamics}

\author{
    \IEEEauthorblockN{Xiaoxing Mo\textsuperscript{*}\IEEEauthorrefmark{1}, Yechao Zhang\textsuperscript{*}\IEEEauthorrefmark{2}, 
    Leo Yu Zhang\textsuperscript{\faEnvelopeO}\IEEEauthorrefmark{3}, 
    Wei Luo\IEEEauthorrefmark{1},
    Nan Sun\IEEEauthorrefmark{4},
    Shengshan Hu\IEEEauthorrefmark{2}, 
    Shang Gao\IEEEauthorrefmark{1},
    Yang Xiang\IEEEauthorrefmark{5}
    }
    \IEEEauthorblockA{\IEEEauthorrefmark{1}Deakin University }
    \IEEEauthorblockA{\IEEEauthorrefmark{2}Huazhong University of Science and Technology}
     \IEEEauthorblockA{\IEEEauthorrefmark{3}Griffith University}
     \IEEEauthorblockA{\IEEEauthorrefmark{4}The University of New South Wales}
     \IEEEauthorblockA{\IEEEauthorrefmark{5}Swinburne University of Technology}
    \IEEEcompsocitemizethanks{
    \IEEEcompsocthanksitem
    \textsuperscript{*} Xiaoxing Mo and Yechao Zhang contributed equally to this work.
    \IEEEcompsocthanksitem 
    \textsuperscript{\faEnvelopeO} Correspondence to Leo Yu Zhang~(\href{mailto:leo.zhang@griffith.edu.au}{leo.zhang@griffith.edu.au}).
    }
}
\maketitle
\thispagestyle{plain}
\pagestyle{plain}

\begin{abstract}
\input{sections/0_abs_1}
\end{abstract}

\section{Introduction}
\label{Sec:intro}

\input{sections/1_intro_1}

\section{Background}
\label{Sec:background}
\input{sections/2_background_1}

\section{Defeating Existing Defenses With Better Attacks}
\label{Sec:SSIDattack}

\input{sections/3_SSID}

\section{\ourdefense 
for Backdoor Detection}
\label{Sec:TEDdefense}
\input{sections/4_TED}

\section{Experimental Analyses}
\label{Sec:exp}

\input{sections/5_Experiment_1}


\section{Discussion, Limitations, and Future Directions}
\label{Sec:discussion}

\input{sections/6_Discussions}



\section{Conclusion}

\input{sections/7_Conclusion}

{
\small
\bibliographystyle{plain}
\bibliography{references}
}

\input{sections/8_appendix}

\end{document}

%% file: sections/0_abs_1.tex
A backdoor attack in deep learning inserts a hidden backdoor in the model to trigger malicious behavior upon specific input patterns. Existing detection approaches assume a metric space (for either the original inputs or their latent representations) in which normal samples and malicious samples are separable. We show that this assumption has a severe limitation by introducing a novel \ourattack (\underline{S}ource-\underline{S}pecific and \underline{D}ynamic-\underline{T}riggers) backdoor, which obscures the difference between normal samples and malicious samples.

To overcome this limitation, we move beyond looking for a perfect metric space that would work for different deep-learning models, and instead resort to more robust topological constructs. We propose \ourdefense (\underline{T}opological \underline{E}volution \underline{D}ynamics) as a model-agnostic basis for robust backdoor detection.
The main idea of \ourdefense is to view a deep-learning model as a dynamical system that evolves inputs to outputs.
In such a dynamical system, a benign input follows a natural evolution trajectory similar to other benign inputs.
In contrast, a malicious sample displays a distinct trajectory, since it starts close to benign samples but eventually shifts towards the neighborhood of attacker-specified target samples to activate the backdoor.

Extensive evaluations are conducted on vision and natural language datasets across different network architectures. 
The results demonstrate that \ourdefense not only achieves a high detection rate, but also significantly outperforms existing state-of-the-art detection approaches,  particularly in addressing the sophisticated \ourattack attack. The code to reproduce the results is made public on {\href{https://github.com/tedbackdoordefense/ted}{GitHub}}.

%% file: sections/1_intro_1.tex
Deep learning with neural networks (DNN) has been shown to be highly effective across various domains, including computer vision, speech recognition, machine translation, and game play~\cite{pouyanfar2018survey}. However, the growing popularity of online machine learning platforms, coupled with the demand for large training sets and extensive computational resources, has led to a new security threat in the DNN model supply chain: the backdoor attack \cite{liu2017trojaning,gu2017badnets,bagdasaryan2021blind,nguyen2020input,li2021invisible,quiring2020backdooring,yang2021careful,gao2019strip,lin2020composite,tang2021demon,jia2022badencoder,carlini2021poisoning,severi2021explanation,adi2018turning,zhong2022attention}. Backdoor attacks involve adversaries creating DNN models with hidden backdoors that can be triggered by specific inputs, potentially leading to disastrous consequences in safety-critical applications, such as autonomous driving and user authentication. Such attacks are difficult to detect because the backdoored models perform well on normal inputs.

Following the seminal work of BadNets \cite{gu2017badnets} in 2017, an influx of backdoor attacks has been proposed in recent years. 
They differ in the level of control the adversary possesses and in the characteristics of the trigger strategy.
By the level of control, backdoor attacks can be classified into: attackers control only the model~\cite{li2021backdoor,liu2017trojaning,yang2021careful}, attackers poison a small part of the training data~\cite{adi2018turning,lin2020composite,tang2021demon,chen2017targeted,zeng2021rethinking}, and attackers control the whole training process~\cite{quiring2020backdooring,nguyen2020input,li2021invisible}. 
Backdoor triggers can be 
static~\cite{gu2017badnets,chen2017targeted,liu2017trojaning,zeng2021rethinking} 
or dynamic \cite{quiring2020backdooring,nguyen2020input,li2021invisible}. 
Triggers can be source-agnostic 
(applied to all classes) \cite{gu2017badnets,li2021invisible,nguyen2020input,yang2021careful} and source-specific (applied to one or few specific classes)~\cite{tang2021demon,lin2020composite}. 
As a general trend,  DNN backdoors become increasingly elusive when attackers implant source-specific and dynamic triggers.

To mitigate this crucial vulnerability of DNNs, many defenses have been proposed, aiming either to remove 
backdoors without impairing normal performance or to detect the existence of backdoors \cite{liu2018fine,wu2021adversarial,tao2022model,huang2022backdoor,li2021anti,xu2021detecting,wang2019neural,liu2019abs,gao2019strip,chou2018sentinet,tang2021demon,ma2022beatrix,liu2023detecting,hu2023pointcrt}. Our focus is backdoor detection since backdoor removal requires model training/re-training and is not suitable for already deployed models. 

For attacks with source-agnostic and static-trigger backdoors~\cite{gu2017badnets,liu2017trojaning}, Neural Cleanse \cite{wang2019neural} synthesizes an artificial trigger pattern that can convert all clean samples to a specified target class. It then separates all synthesized artificial trigger patterns by checking their abnormality, measured in the $l_1$ distance. 
STRIP \cite{gao2019strip} superimposes a test sample with a set of randomly selected benign samples to collect a set of confidence score vectors. It then separates benign and  samples by examining the distribution difference between two sets of confidence vectors.
%
To thwart source-specific attacks~\cite{tang2021demon,lin2020composite}, SCAn performs a two-component decomposition on the latent representations of images with the EM algorithm, and then separates benign and malicious images according to a weighted Mahalanobis distance~\cite{tang2021demon}. 
In the most recent effort to resist dynamic-trigger backdoors, Beatrix~\cite{ma2022beatrix} uses the Gramian matrix to capture both the latent features' correlation and their high-order information, and then separates benign and malicious images according to the median absolute deviation.

\begin{table*}[t]
\centering
\caption{\label{tab:Defense Solutions} Comparison of different backdoor defenses ($\blacksquare$ and $\square$ denote the defense supports this property or not).}
\begin{threeparttable}[b]
\begin{adjustbox}{width=0.9\textwidth,center}
\begin{tabular}{@{}lllllllllll@{}}
\toprule
\multirow{2}{*}{Analysis}             & \multirow{2}{*}{Backdoor} & \multirow{2}{*}{No  Clean} & \multicolumn{3}{l}{Detection Level}                                       & \multicolumn{5}{l}{Trigger Strategy}                                                                                 \\ 
\cmidrule(l){4-6} \cmidrule(l){7-11} 
                                   {Method}     & Detector                           &  Data                                      & \multirow{2}{*}{Sample}        & \multirow{2}{*}{Label}        & \multirow{2}{*}{Model} & {Source} & {Source}     & {Static} & {Dynamic}       & 
                                   {\ourattack}  \\
                                      &                            &                                        &                                &                               &                        & Agnostic       & Specific       & Trigger         & Trigger        & (new strategy)               \\ \midrule  
\multirow{2}{*}{Model   Meta}         & MNTD~\cite{xu2021detecting}                     & $\square$                              & $\square$                      & $\square$                     & $\blacksquare$         & $\blacksquare$ & $\square$      & $\blacksquare$ & $\square$      & $\square$                                          \\
                                      & ABS~\cite{liu2019abs}                       & $\square$                              & $\square$                      & $\blacksquare$                & $\square$              & $\blacksquare$ & $\square$      & $\blacksquare$ & $\square$      & $\square$                                          \\ \midrule
\multirow{3}{*}{Input   Perturbation} & Neural   Cleanse~\cite{wang2019neural}         & $\square$                              & $\square$                      & $\blacksquare$                & $\square$              & $\blacksquare$ & $\square$      & $\blacksquare$ & $\square$      & $\square$                                        \\
                                      & STRIP~\cite{gao2019strip}                    & $\square$                              & $\blacksquare$                 & $\square$                     & $\square$              & $\blacksquare$ & $\square$      & $\blacksquare$ & $\square$      & $\square$                     \\
                                      & SentiNet~\cite{chou2018sentinet}                   & $\square$                              & $\blacksquare$                 & $\square$                     & $\square$              & $\blacksquare$ & $\square$      & $\blacksquare$ & $\square$      & $\square$                                       \\ \midrule
\multirow{4}{*}{Latent Feature}              & Activation-Clustering~\cite{chen2018detecting}      & $\blacksquare$                         & $\blacksquare$                 & $\square$                     & $\square$              & $\blacksquare$ & $\square$      & $\blacksquare$ & $\square$      & $\square$                                        \\
                                      & Spectral-Signature~\cite{tran2018spectral}        & $\blacksquare$                         & $\blacksquare$                 & $\square$                     & $\square$              & $\blacksquare$ & $\square$      & $\blacksquare$ & $\square$      & $\square$                                        \\
                                      & SCAn~\cite{tang2021demon}                       & $\square$                              & $\blacksquare$ & $\blacksquare$ & $\square$              & $\blacksquare$ & $\blacksquare$ & $\blacksquare$ & $\square$ & $\square$                                        \\
                                      & Beatrix~\cite{ma2022beatrix}                    & $\square$                              & $\blacksquare$ & $\blacksquare$ & $\square$              & $\blacksquare$ & $\blacksquare$ & $\blacksquare$ & $\blacksquare$ & $\square$                                         \\ \midrule
Topology   Analysis                  & \ourdefense   (our work)           & $\square$                              & $\blacksquare$                 & $\square$                     & $\square$              & $\blacksquare$ & $\blacksquare$ & $\blacksquare$ & $\blacksquare$ & $\blacksquare$          \\ \bottomrule
\end{tabular}
\end{adjustbox}
\end{threeparttable}
\vspace{-1mm}
\end{table*}

{Prior studies, however, either implicitly~\cite{gao2019strip,wang2019neural} or explicitly \cite{tang2021demon,ma2022beatrix}  assume that with appropriate pre-processing of the raw samples or their latent features, benign and malicious samples can be separated under certain metrics in the \textit{metric space}. To assess the worst-case security against backdoors, we propose \ourattack, which obscures the difference in features between these two types of samples with both \underline{S}ource-\underline{S}pecific and \underline{D}ynamic-\underline{T}rigger strategies under the strongest knowledge of attackers (i.e., control the whole training process). Under this new attack, existing defenses fail to effectively separate benign and malicious samples~(see Sec.~\ref{sec:limitations}).}

{This observation motivates us to switch the defense rationale from the \textit{metric space} to the more general \textit{topological space}, which focuses on studying the neighborhood relationship for each sample based on the concept of closeness (\eg, close in distance). This alternative view enables us to capture the root difference between benign and malicious samples. In essence, for any well-trained DNN model, a benign sample (with label $y$) will likely be surrounded by a large number of neighboring samples from the same class as it propagates deeper into the network. Conversely, a malicious sample (adapted from the source label $y$ but aimed at the target label $t$) typically remains close to samples drawn from label $y$ initially and then moves closer to samples drawn from label $t$ as it propagates. By capturing this root difference in topological structures as a feature, 
we design \ourdefense (\underline{T}opological \underline{E}volutionary \underline{D}ynamics), a novel backdoor detector that can operate with simple outlier-detection methods such as PCA (principal component analysis). An overall comparison of \ourdefense with the SOTA (state-of-the-art) backdoor detectors is presented in Table~\ref{tab:Defense Solutions}. }

\noindent\textbf{Contributions.} The contributions of this work are twofold.
\begin{itemize}[leftmargin=*]
\item \textbf{{New understanding}}: 
We carefully review and classify trigger strategies that appeared in SOTA backdoor attacks. The analysis reveals an immediate drawback of existing backdoor detectors, which aim to differentiate benign samples from malicious samples by separating their raw/latent features in the metric space. We show that \ourattack, which obscures the features of benign and trigger-carrying samples with both source-specific and dynamic-trigger strategies, can invalidate all SOTA detectors. 

\item \textbf{{New detection}}: 
We shift the underlying defense rationale from metric space to topological space and propose a new detection method called \ourdefense. It extracts features from topological structures when a sample propagates in the network. Extensive experimental results across different network architectures and datasets demonstrate that \ourdefense significantly outperforms existing SOTA detectors, particularly in addressing the sophisticated \ourattack attack.
\end{itemize}

%% file: sections/2_background_1.tex
\subsection{Deep Neural Backdoor Attacks}
\label{Subsec:DNNBack}
\noindent\textbf{Deep Neural Network.} A deep neural network model $f$ comprises multiple layers $\{f_{n}: n \in [1, N]\}$, with each layer being a
transformation function.
Given an input $x$, the output of the neural network $f$ is computed as 
$$f(x) = \left(f_{N} \circ \cdots \circ f_1 \right)(x).$$
Following~\cite{gu2017badnets,bagdasaryan2021blind,nguyen2020input,li2021invisible,yang2021careful,gao2019strip,lin2020composite,tang2021demon}, this paper focuses on DNN models for classification. 
Thus, the network can be further decomposed into two parts, $f_{N-1}\circ \cdots \circ f_1$ and $f_N$, where the former extracts the representation of sample $x$, and $f_N$ is the classifier based on the output of penultimate layer $f_{N-1}$.
Specifically, we consider a $c$-classes classification problem with normal input space  $X$ and label space $Y=\{1, ..., c\}$. 
Any ground-truth input-output pair $(x, y)$  lies in the normal distribution $\mathcal{P}_{x, y}$, which is supported on  $(X, Y)$.
We can further denote the marginal input space that comprises all the inputs whose ground-truth labels are all $j$ as $X_j = \{x|(x,y)\sim \mathcal{P}_{x, 
y=j} \}$.

Given a standard classification training dataset $D = \left\{ (x_i, y_i) \right\}$ consisting of data points $(x_i, y_i) $ drawn from $\mathcal{P}_{x, y} $,
$f$ is trained by minimizing a loss function $L(\cdot, \cdot)$,
\ie, $ f^* = \argmin_{f} \sum_i L\left(y_i, f(x_i) \right)$. 
Once trained, model $f$ outputs a confidence score vector $f(x)$ for any test sample $x$, and takes $\argmax_{k \in [1, c]}  f(x)_k$ as the classification result. \\

\noindent \textbf{Backdoor Attacks.} Backdoor attacks on neural networks involve an attacker embedding a malicious functionality into a neural network model. 
Such a compromised model behaves normally when processing normal inputs, but could present malicious behavior when presented with trigger-carry samples. Notably, a trigger-carrying sample may not always activate the backdoor, particularly if it targets an unintended class.
There are a plethora of new attacks that have been proposed in recent years, covering different applications like the classification of images or texts \cite{gu2017badnets,bagdasaryan2021blind,nguyen2020input,li2021invisible,yang2021careful,lin2020composite,tang2021demon}, semi-/self-supervised learning \cite{jia2022badencoder,carlini2021poisoning}, malware detection \cite{severi2021explanation}, ownership verification \cite{adi2018turning,zhong2022attention}, and more. 
Backdoors in classification tasks can be classified according to the type of trigger and the victim class(es) affected by the backdoor. 
We elaborate on this in the following sections.
\subsubsection{Static-Trigger and Dynamic-Trigger}
\label{subsubsec:StaticorDynamic}
Considering the trigger patterns utilized for backdoor implants, backdoors can be classified into two categories: static-trigger and dynamic-trigger.

\noindent\textbf{Static-trigger backdoor.} 
In a static-trigger backdoor attack, though the trigger can be in various forms \cite{gu2017badnets,chen2017targeted,liu2017trojaning,zeng2021rethinking}, all corrupted samples share the same trigger. 
The process of generating corrupted samples can be mathematically represented as a function $A_{ST}: x \mapsto A_{ST}(x)$, where $x$ is a clean sample, and $A_{ST}(x)$ is the corrupted sample. 
This can be defined formally as
$$
A_{ST}(x)  =   x \oplus  \delta,
\label{Eq:staticTrigger}
$$
where 
$\delta$ is a fixed trigger pattern and $\oplus$ represents the general   operation of superimposing $\delta$ to $x$.
The corrupted samples generated by $A_{ST}(\cdot)$ are then labeled as the attacker's target $t$ to get the backdoor dataset $D_b =\{A_{ST}(x), t\}$.
The malicious functionality (\ie, classifying test samples stamped with a trigger as $t$ regardless of the samples' semantics) will be embedded into the model $f$ if $D \cup D_b$ is used for training \cite{gu2017badnets,chen2017targeted, liu2017trojaning} or model fine-tuning \cite{zeng2021rethinking}.  

\noindent\textbf{Dynamic-trigger backdoor.} On the contrary, dynamic-trigger backdoor attacks do not utilize a fixed pattern for the trigger across all corrupted samples.
Instead, the trigger varies from input to input, making the attack more stealthy and challenging to detect and defend against. 
Denote $g$ as a pattern generator, the process of generating dynamic corrupted samples $A_{DT}$ can be formulated similarly, \ie, 
\begin{IEEEeqnarray}{rCL}
A_{DT}(x) &=&  x  \oplus  g(x).
\label{Eq:DynamicTrigger}
\end{IEEEeqnarray}
In contrast to the static-trigger backdoor, the trigger pattern $g(x)$ is conditioned to each input $x$.
Generally, the generator could be a parameter-free design \cite{quiring2020backdooring} or a network module with learnable parameters \cite{nguyen2020input,li2021invisible,hu2022badhash}.
For example, \cite{quiring2020backdooring} employed an image scaling operation as $g$, and \cite{li2021invisible,nguyen2020input} instantiated $g$ with an encoder-decoder architecture. 
To create stronger attacks with more diverse and adaptive patterns, the parameterized version of $g$ can be co-optimized together with the to-be-backdoored model $f$, \ie, 
$g^*, f^* = \argmin_{g,f} \left( \sum\limits_{x_i \in D} L\left(y_i, f(x_i) \right) + \sum\limits_{x\in D_b} L\left(t, f(A_{DT}(x)) \right) \right)$, where $D_b = \{A_{DT}(x), t\}$ is the backdoor dataset associated with  the target label $t$.

\subsubsection{Source-Agnostic and Source-Specific}
\label{Subsecsec:SourceAorS}
{
Parallel to the categorization above, considering the victim source class(es) that the trigger-carrying sample targets, backdoors can also be classified as source-agnostic and source-specific types.
For ease of presentation, our illustration below focuses on the case of one target label only. }



\noindent\textbf{Source-agnostic attack.} 
In a source-agnostic backdoor attack, irrespective of the original class, any input with the trigger is misclassified to the target label $t$ by the infected model \cite{gu2017badnets,li2021invisible,nguyen2020input,yang2021careful}. 
For any test sample $x$ with its ground-truth label $y$, a source-agnostic backdoored classifier behaves as
\begin{IEEEeqnarray}{rCl} 
\argmax_k f(x)_k &= &y, \\ 
\argmax_k f(A(x))_k &= &t,
\IEEEstrut 
\end{IEEEeqnarray}
{where the trigger function $A$ could be $ A_{ST}$ or  $A_{DT}$ in the literature.  
Such a source-agnostic backdoor makes the representation of any stamped input predominantly affected by the trigger. 
As a result, it tends to behave quite differently from that of a normal input with the target label~\cite{tang2021demon}, making it hard to bypass sophisticated detection strategies.}



\noindent\textbf{Source-specific backdoor.} 
On the other hand, a source-specific backdoor attack affects a  chosen victim source class $s$. 
That is, inputs from all non-victim classes, even when stamped with the trigger, will not be misclassified into the target label $t$. 
For any test sample $x$ with its  ground-truth label $y$, a source-specific backdoored classifier should behave as 
\begin{IEEEeqnarray}{rCl} 
\argmax_k f(x)_k &= &y, \\ 
\argmax_k f(A(x))_k &= &t, \text{\quad if }  x \in X_s,\\
\argmax_k f(A(x))_k &= &y, \text{\quad if }  x \notin  X_s.
\IEEEstrut 
\end{IEEEeqnarray}

The source-specific backdoor implantation process utilizes the backdoor dataset $D_b$ \cite{tang2021demon,lin2020composite}, in conjunction with the \textit{laundry dataset} $D_l$, which are constructed as follows:
\begin{IEEEeqnarray}{rCl} 
&D_b = \{(A(x), t) \mid   (x,y) \in D, x \in X_s  \},  \label{Eq:backdoor dataset} \\ 
&D_l = \{(A(x), y) \mid  (x,y) \in D, x \notin X_s  \label{Eq:laundry dataset} \}.
\end{IEEEeqnarray}
The purpose of the laundry dataset $D_l$ is to conceal the malicious backdoor effect from $D_b$ by implanting a conditional trigger. Specifically, the backdoor is activated only in the presence of both the trigger pattern and samples from the victim class.
Such a conditional design enforces the backdoored model to classify the malicious samples in a way strongly dependent on the features used in the normal classification, thus dispersing the predominant effect of the trigger and making samples in $D_l$ \textit{indistinguishable} from those of normal samples.
It should be noted that the current source-specific attacks follow the static trigger routine with  $D_b$ and $D_l$ sharing a fixed trigger {$\delta$~\cite{tang2021demon}}.
In Section \ref{Sec:SSIDattack}, we introduce a stronger attack paradigm that includes both the source-specific attack goal and the dynamic trigger design to serve as the strongest evaluation baseline.



\subsection{Existing Backdoor Detection Methods}

At a high level, existing detection methods can be broadly categorized based on their focus objectives into three types: \textit{model-level}, \textit{label-level}, and \textit{input-level} detections.
At the \textit{model-level}, the defense's objective is to ascertain whether a model is compromised by a backdoor.
Under a black-box access limitation,  Meta Neural Trojan Detection (MNTD)~\cite{xu2021detecting} trains a meta-classifier based on a set of backdoored models following a general distribution.

With white-box access, more granular detection is achievable. 
In the context of \textit{label-level} detection, not only the backdoored model itself can be detected, but also the infected label where the backdoor targets can be identified.
Such defenses include Neural Cleanse~\cite{wang2019neural} and ABS~\cite{liu2019abs}. 
Neural Cleanse considers the label whose reversed potential trigger pattern exhibits the minimum norm
as infected.
Meanwhile, ABS believes that the neuron activations for the infected label are elevated. 
However, identifying the infected label does not determine whether samples classified into it are normal or malicious, which still does not provide the required protection for model consumers. 

\textit{Input-level} detection, on the other hand, separating the malicious samples from the clean ones, provides the most granular level of detection.
Existing detection methods in this category, including STRIP~\cite{gao2019strip}, SentiNet~\cite{chou2018sentinet}, SCAn~\cite{tang2021demon}, and Beatrix~\cite{ma2022beatrix}, fundamentally rely on the feature separability between normal and malicious samples in some metric space (e.g., Euclidian space).
Some of them hold the assumption that the feature of the trigger pattern is independent of the normal feature, thus dominating the prediction of backdoored models with consistently low entropy~\cite{gao2019strip} and high confidence~\cite{chou2018sentinet} when trigger patterns are presented.
However, as we will detail in Section \ref{Sec:SSIDattack}, when encountering stronger attacks that make the representation of malicious inputs indistinguishable from normal inputs, features separability in certain metric space is no longer valid and defenses are doomed to fail.

\section{Problem Statement}
Here, we provide an overview of how we explore and address the limitations of current input-level defense methods and present our solutions in this paper. 
Specifically, we start by considering an adversary equipped with maximum capabilities and proceed to design the \ourattack attack, which combines all the hard-to-detect properties to benefit the adversary (Sec. \ref{sec:ourattack}). 
Subsequently, we demonstrate how all existing defense strategies fail to defend against \ourattack, as their shared assumption of feature separability in the metric space is violated under such a strong adversary (Sec. \ref{sec:limitations}). 
Confronting these challenges, we propose a new defense approach, \ourdefense, which leverages the evolution dynamic of topological structures throughout the network and does not rely on the compromised assumption (Sec. \ref{Sec:TEDdefense}).
Finally, a comprehensive evaluation of \ourdefense (Sec.~\ref{Sec:exp}) underscores its efficacy in countering various backdoor attacks, including \ourattack.

\subsection{Threat Model of \ourattack}\label{subsec:threat-ssdt}
\noindent 
{\textbf{Attack goals.} The adversary aims to implant a backdoor in models provided to consumers. The backdoored model misclassifies inputs with a particular trigger associated with selected class(es) to a predefined target label, while remaining accurate for other inputs.
In the design of \ourattack,  the intent is to challenge and reveal the shortcomings of current \textit{input-level} defense strategies. 
}

\noindent 
\textbf{Adversary's capabilities and knowledge.} 
To better explore the limitations of existing defense strategies, we aim to maximize the capabilities of \ourattack adversary.
Thus, we assume that the adversary possesses complete control over certain data sources, allowing them to manipulate the data at their discretion. 
Moreover, we assume the \ourattack adversary has sufficient knowledge of existing defense strategies,  thus enabling the creation of a general approach that circumvents all known defense strategies.
Also, we consider the scenarios where the adversary might be aware of the existence of the \ourdefense and strategies to evade the defense {(Sec. \ref{sec:adaptive-attack})}. 
The strong assumptions about the adversary's capabilities make it possible to design stronger backdoor attacks, creating the worst-case condition for defenders.

\subsection{Defense Assumptions of \ourdefense}
\noindent{\textbf{Defense goals.} The objective of the defender is to develop an effective \textit{input-level} defense strategy, \ie, 
determining whether a given model is backdoored from the instances it classifies and identifying those malicious samples. 
Such input-level detection provides the most granular level of defense by evaluating individual inputs for malicious activities.
The other purpose of \ourdefense is to surpass the limitations of existing \textit{input-level} defense strategies. 
}

\noindent {\textbf{Defender's capabilities and knowledge.}  We consider the defender has full access to the given model, including all the outputs in each intermediate layer, but cannot interfere with the model's training process. We also assume the defender possesses a small amount of clean data to build its detection strategy. Further, we assume the \ourdefense defender is blind to any potential attack strategy, \ie, what kind of attack is deployed, and even whether the model is backdoored or not is unknown to the defender.
The limited knowledge further enlarges the difficulty for defenders and necessitates a once-for-all strategy for any potential backdoor attacks.
}

\subsection{Datasets, Models, and Metrics}
\label{subsec:DataModel}
In both evaluations for \ourattack and \ourdefense, we utilize three datasets to evaluate the performance and robustness.

\textit{MNIST}~\cite{lecun1995learning}. This dataset is a standard benchmark in handwritten digit recognition. It comprises grayscale images of digits (0 to 9), making up a total of 70,000 images—60,000 for training and 10,000 for testing. We utilize MNIST due to its simplistic data structures, allowing us to evaluate our proposed defense method's performance in a less complex scenario. For this dataset, \ourattack incorporates a Convolutional Neural Network (CNN) model with two convolutional layers and two fully connected layers. 

\textit{CIFAR-10} \cite{CIFAR10}. This more complex dataset has 60,000 color images of size 32$\times$32 across ten classes.  Given its high intra-class variability and intricate patterns, CIFAR-10 introduces an elevated level of complexity. \ourattack leverages PreAct-ResNet18 \cite{he2016identity} as the target model, in line with the approach taken by Nguyen et al.~\cite{nguyen2020input}, to evaluate its performance in more complex scenarios.

\textit{GTSRB}~\cite{stallkamp2012man}. This dataset presents real-world challenges due to the variability in the size and shape of traffic sign images. Using this dataset, we evaluate the robustness of \ourattack under diverse and constrained conditions, with PreAct-ResNet18 being the target model.

{We assess the performance of backdoored models in terms of their classification accuracy under different data types,  including no-trigger samples (Acc NoT, also known as clean accuracy), victim-triggered samples (Acc VT, also known as attack success rate), non-victim but triggered samples (Acc NVT), and cross-triggered samples (Acc CT).}
%
We evaluate the performance of backdoor detectors by computing the True Positive Rate (TPR) and False Positive Rate (FPR). These metrics appraise the defense's sensitivity and specificity. 
The Area Under ROC Curve (AUC) for \ourdefense is calculated when performing the ablation study ({Sec.~\ref{subsec:AblationStudy}}). 

\begin{figure*}[!htbp] 
\centering
  \centering
    \subcaptionbox{Clean task}{\includegraphics[width=0.147 \textwidth]{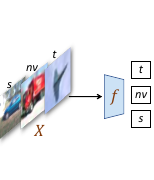}}
    \hfill
      \subcaptionbox{Backdoor task}{\includegraphics[width=0.213\textwidth]{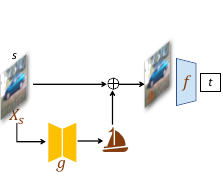}}
      \hfill 
      \subcaptionbox{Laundry task}{\includegraphics[width=0.268\textwidth]{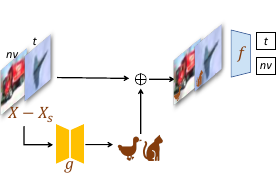}}
    \hfill 
    \subcaptionbox{Cross task}{\includegraphics[width=0.308\textwidth]{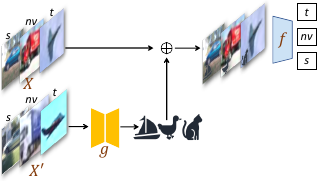}}
      
      \caption{\ourattack training tasks: Clean,  Backdoor, Laundry,  and Cross. }
\label{fig:adv-transfer}
\end{figure*}



%% file: sections/3_SSID.tex
\begin{algorithm}[t]
 Given a target label $t$, a clean training set $D$, 
 the source victim class samples set $X_{D_s}$, the non-victim class set $X_{D_{nv}}$, clean probability $\rho$, backdoor probability $\rho_b$, laundry probability $\rho_l$, cross trigger probability $\rho_{ct}$ 
 
  \textbf{{Initialize}}
  $f$ and $g$\;
  \For{\textit{number of iterations}}
  {

  $d \leftarrow \mathtt{random}(0,1)$\;
   \uIf{$d < \rho$ }
   {
   $(x, y) \leftarrow \mathtt{random\_sample}(D)$;
    $L_{min} \leftarrow L(y, f(x))$
  }
  \uElseIf{$d < \rho + \rho_b$}{
  $x  \leftarrow \mathtt{random\_sample}(X_{D_s})$
     $L_{min} \leftarrow L(t, f(x \oplus g(x))) $
  }
  \uElseIf{$d < \rho + \rho_b + \rho_l$}{
   $(x, y) \leftarrow \mathtt{random\_sample}(D_{nv})$
   $L_{min} \leftarrow L(y, f(x \oplus g(x))) $
  }
  \Else{
   $(x, y), (x', y') \leftarrow \mathtt{random\_two\_samples}(D)$
   $L_{min} \leftarrow L(y, f(x \oplus g(x'))) $
  }
  $g, f \leftarrow \mathtt{optimize}(L_{min})$
  
}

  \Return $g, f$\;

  \caption{\ourattack}

  \label{algo:ssid}

  \vspace{-1mm}
\end{algorithm}

\subsection{Formulation of \ourattack}
\label{sec:ourattack}
As mentioned earlier, we introduce 
a new attack paradigm \ourattack which blends source-specific and dynamic-trigger backdoors with more advanced attack goals.
To further clarify this intuition,
here we formulate the attack goals of  \ourattack.
Denote $X_s$  as the source victim class and  $t$ as the target label, then for any two different clean data points  $(x,y), (x', y')\sim \mathcal{P}_{x,y}$ 
 ($y=y'$or $y\neq y'$), \ourattack should behave as
\begin{IEEEeqnarray}{rCl}
 \underset{k}{\argmax}~f(x)_k=y,& \label{goal1} \\
 \underset{k}{\argmax}~f( x \oplus g(x)  )_k=t,& \quad \text {if } x \in X_s,  \label{goal2} \\
 \underset{k}{\argmax}~f( x \oplus g(x)  )_k=y,& \quad \text {if } x \notin X_s,  \label{goal3} \\
\underset{k}{\argmax}~f( x \oplus g(x')  )_k=y.& \label{goal4}
\end{IEEEeqnarray}
Note that Eqs. \ref{goal1}, \ref{goal2} and \ref{goal3} depict the goals for the source-specific attack as we mentioned in Sec.~\ref{Subsecsec:SourceAorS}, and the learnable $g$ produces the dynamic triggers.
In addition, referring to \cite{nguyen2020input}, here we add another requirement, as illustrated by Eq. \ref{goal4}, that the trigger generated from another sample $x'$ cannot change the prediction for $x$, to ensure the non-reusability of the trigger. 
This alone will force $g$ to produce more diverse trigger patterns for different inputs.
Further, combining different requirements makes the ultimate attack {goal} even stricter.
\textit{First}, the joint goal of Eqs. \ref{goal2} and \ref{goal4}  suggest  even when both $x, x' \in X_s$, the backdoor activation of $x$ can only be obtained by its own specific trigger pattern $g(x)$.
\textit{Second}, the joint goal of Eqs. \ref{goal3} and \ref{goal4}  suggest the representations of non-victim class samples are robust to any trigger pattern, whether it comes from their own or not.

In practice, we co-optimize $g$  with $f$, 
and the exact data points used for training the model are generated along with the optimization of $g$.
To this end, we prepare the clean training set $D$, the source victim class samples  $X_{D_s}=\{x| (x, y) \in D, x \in X_s\}$, and non-victim class dataset $D_{nv} = \{(x, y)| (x, y) \in D, x \notin X_s \}$ beforehand, and always utilize the updated $g$ to produce the trigger-carrying samples in each optimization iteration.
During training, we treat the optimizations of  Eqs. \ref{goal1}-\ref{goal4} respectively as four tasks, \ie, clean task, backdoor task, laundry task, and cross task.
In each iteration, we randomly chose an optimization task, with the probabilities of these four tasks as clean probability $\rho$, backdoor probability $\rho_b$, laundry probability $\rho_l$, and cross trigger probability $\rho_{ct}$, respectively, where $\rho + \rho_b + \rho_l + \rho_{ct} =1$.
Algorithm~\ref{algo:ssid} illustrates the training process  {and Fig.~\ref{fig:adv-transfer} visualizes how different types of data are prepared.} 

\begin{table}[h]
\centering
\caption{\label{tab:SSID and Benign ACC} Accuracy (\%) for \ourattack and benign models.}
\begin{tabular}{@{}llllll@{}}
\toprule
\multirow{2}{*}{Dataset} & \multicolumn{4}{c}{\ourattack}                                                    & Benign         \\ 
\cmidrule(lr){2-5} \cmidrule{6-6}
                         & NoT & VT & NVT & CT & Clean \\ 
                         \midrule
MNIST                    & 98.98        & 99.47             & 97.18                  & 97.03   & 99.37        \\ 
CIFAR-10                  & 93.68        & 98.40             & 93.39                  & 89.63   & 94.50        \\
GTSRB                    & 98.86        & 99.31             & 97.57                  & 94.98   & 99.10        
\\
\bottomrule
\end{tabular}
\end{table}


We evaluate the attack performance of \ourattack based on its four optimization goals (Eqs. 9-12) and the results are reported in Table \ref{tab:SSID and Benign ACC}.
The accuracy of no-trigger samples (NoT),
the accuracy of victim-triggered samples (VT),
the accuracy of non-victim but triggered samples (NVT), and the accuracy of cross-triggered samples (CT) correspond to the Eqs. \ref{goal1}-\ref{goal4}, respectively. 
Further, we also train a benign model with the same training setting for comparison.
As we can see, the results on all the three datasets demonstrate the effectiveness of the attack. 
\ourattack yields significant attack performance for victim class samples while the clean accuracy is slightly lower than the benign model. 
On the other hand, the non-victim class samples with triggers 
(NVT) have accuracy comparable to the clean data under the \ourattack backdoored model.

\subsection{Limitations of Existing Defense Solutions Against \ourattack}
\label{sec:limitations}

\begin{figure}[h]
  \centering
  \includegraphics[width=0.2\textwidth]{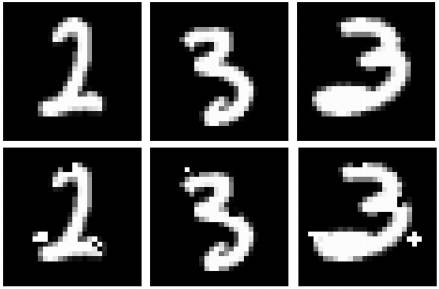}
 \caption{Examples of normal (first row) and dynamic-trigger (second row) samples  on MNIST.}
   \label{fig:Input_Dynamic_Samples}
\end{figure}

In this section, we evaluate four SOTA input-level detection methods against \ourattack: STRIP~\cite{gao2019strip}, SentiNet~\cite{chou2018sentinet}, SCAn~\cite{tang2021demon}, and Beatrix~\cite{ma2022beatrix}. 
Our evaluations are conducted on three datasets: MNIST, CIFAR-10, and GTSRB. For each dataset, we train both benign and \ourattack-backdoored models. 
We follow the input-dynamic module in \cite{nguyen2020input} to take an encoder-decoder architecture as $g$ for implementing \ourattack. 
When implementing the \ourattack within a model, for each designated target label, samples from the subsequent class—or from class 0 if the target label is the highest—are assigned as victims.
Fig.~\ref{fig:Input_Dynamic_Samples} depicts several trigger-carrying samples from MNIST. 


To evaluate backdoor detectors, we {randomly select 1000 images as the training set from the clean dataset for each detector}. For a fair comparison, each detector's threshold is determined by setting {FRP to 5\% for NoT samples} during the training. The test set consists of 4000 randomly selected images---half benign and half victim-triggered---with benign samples equally split into no-trigger and non-victim-triggered categories. The test TPR and FPR are reported in Table~\ref{tab:FPRs and TPRs of defenses on MNIST, CIFAR-10, and GTSRB under SSDT attacks}.

\begin{figure}[h]
  \centering
  \subfloat[VT v.s. NoT
  ]{\includegraphics[width=0.23\textwidth]{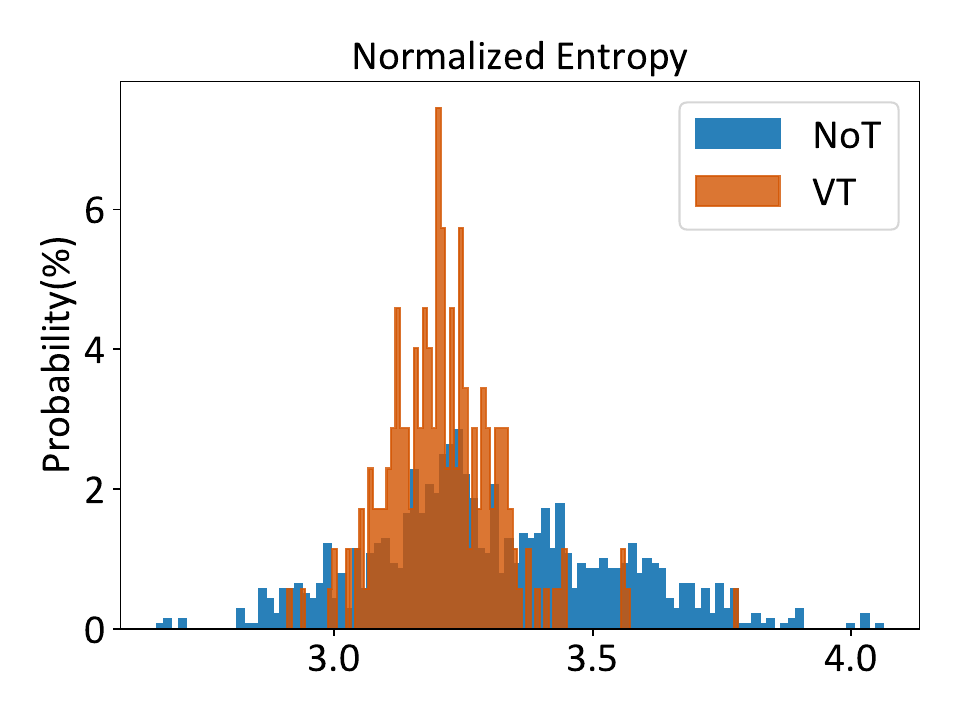}\label{fig:strip1}}
  \subfloat[NVT v.s. NoT]{\includegraphics[width=0.23\textwidth]{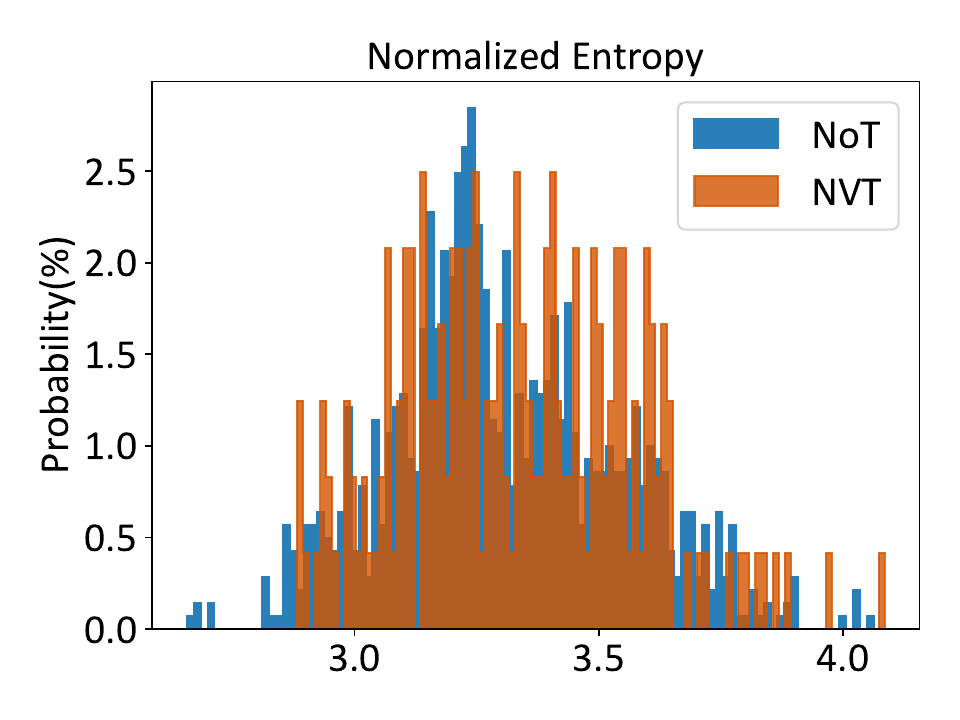}\label{fig:strip2}}
  \caption{Entroy histograms of STRIP.}
  \label{fig:strip}
  \vspace{-1mm}
\end{figure}

\noindent\textbf{SentiNet.} SentiNet \cite{chou2018sentinet} works by separating salient latent features, which strongly affect the model's decision due to the use of a localized static trigger, between benign and malicious samples. 
However, it faces considerable challenges when combating  \ourattack, which involves non-localized and dynamic-trigger. 
These dynamically crafted triggers for each sample heighten the unpredictability of the attack, disrupting SentiNet's capacity to localize and identify common adversarial features.
Moreover, these dynamic triggers are non-reusable, contradicting SentiNet's assumption that triggers are static.

The practical implications of these limitations are exemplified by our empirical results. SentiNet's detection efficacy against \ourattack is significantly reduced, with TPR for VT samples ranging between 2.75\%, 6.75\% and 5.25\% on MNIST, CIFAR-10 and GTSRB, respectively.\\

\noindent\textbf{STRIP.}  STRIP~\cite{gao2019strip} attempts to counteract backdoor attacks by examining 
whether overlaying the input image with a set of randomly selected images obscures the classification results, quantified by the entropy of the logits output of images. 
If the classification is obscured (\ie, high entropy), the input is deemed normal; otherwise, it may contain a trigger. 
Apparently, this approach is based on the assumption that the trigger will significantly affect an image's representation. The hypothesis is that the presence of a trigger will definitely dominate the output of penultimate layer $f_{N-1}$, thus affecting the prediction on $f_N$ to such an extent that even a random image content containing the trigger will be classified as the target label.

For source-specific attacks like \ourattack, however, the influence of the trigger is not as dominant. 
A malicious input's representation also hinges on the characteristics of its source label (the original label of the input).
Since overlaying combines features of two images, it weakens the trigger's connection with the source label and further reduces its power to mislead classification. This diminishes the effectiveness of STRIP.

Our research assesses the performance of STRIP against \ourattack attacks. We generate the logit output for two types of images using models infected by \ourattack on MNIST: those overlaying malicious images onto normal ones and those overlaying normal images onto normal ones. Our results, as illustrated in Fig. \ref{fig:strip}a, demonstrate a critical limitation of STRIP - the difficulty it faces in distinguishing between the entropy distributions of the VT and NoT images. This limitation is due to the overlap between these two distributions.

Furthermore, STRIP's effectiveness to detect source-specific attacks is also affected by its randomness in image selection across all classes for superimposing an input \cite{gao2019strip,tang2021demon}. This means that the chances of detecting an attack input may increase when a large number of images from the source of the attack are chosen to evaluate the input (from the same source and containing a trigger). To investigate this, we conduct an experiment where only the benign images from the source victim class of \ourattack are utilized for determining the predictability of the input, giving STRIP a huge advantage that the victim class is already known. As Fig.~\ref{fig:strip}b indicates, STRIP fails to distinguish them in \ourattack. We believe this is due to the dynamic trigger design and the non-reusability assurance of Eq. \ref{goal4}, which makes the trigger conditional to the specific input, superimposing the trigger-carrying source victim class sample on any sample, even from the same class,  also violates the non-reusability.







\begin{table}[]
\centering
\caption{\label{tab:FPRs and TPRs of defenses on MNIST, CIFAR-10, and GTSRB under SSDT attacks} 
{FPRs and TPRs (\%) of different detectors on MNIST, CIFAR-10, and GTSRB under \ourattack}.}
\begin{tabular}{ccccccc}
\toprule
\textbf{}                     & \textbf{} & \textbf{\ourdefense} & \textbf{Beatrix} & \textbf{SCAn} & \textbf{STRIP} & \textbf{SentiNet} \\ \midrule
\multicolumn{7}{c}{MNIST}                                                                                                        \\ 
\textbf{TPR}                  & VT        & 100.00     & 82.50          & 44.00       & 0.50         & 2.75            \\ \cmidrule(l){3-7}
\multirow{2}{*}{\textbf{FPR}} & NVT       & 1.30       & 4.27           & 5.00        & 4.00         & 4.50            \\

                              & NoT       & 5.00       & 4.55           & 5.00        & 5.50         & 4.50            \\ \midrule 
\multicolumn{7}{c}{CIFAR-10}                                                                                                      \\ 
\textbf{TPR}                  & VT        & 100.00     & 90.00          & 36.50       & 0.00         & 6.75            \\ \cmidrule(l){3-7}
\multirow{2}{*}{\textbf{FPR}} & NVT       & 4.00       & 46.65           & 5.00        & 7.00         & 5.00            \\
                              & NoT       & 5.50       & 4.15           & 5.00        & 6.00         & 4.50       \\ \midrule
                              \multicolumn{7}{c}{GTSRB}                                                                                   \\ 
\textbf{TPR}                  & VT        & 100.00     & 100.00         & 99.50       & 0.50         & 5.25            \\ \cmidrule(l){3-7}
\multirow{2}{*}{\textbf{FPR}} & NVT       & 0.90       & 62.55          & 5.00        & 4.00         & 5.00            \\
                              & NoT       & 4.59       & 4.10           & 5.00        & 5.50         & 5.50           \\ 
                              \bottomrule
\end{tabular}
\vspace{-1mm}
\end{table}

\noindent\textbf{SCAn.} 
SCAn \cite{tang2021demon} conducts decomposition on the latent representations of images with the EM algorithm and separates benign and malicious images by the first-moment (mean) discrepancy. 
In particular, it assumes that a benign image's representation vector $r_y$ can be described as the sum of two latent vectors, class-wise identity vector $\mu_y$ and universal variation vector $e$, each following a normal distribution, $\mu_y \sim N(0; S_{\mu_y})$ and $e \sim N(0; S_e)$. However, the representation of a trigger image (with source label $y$ but targeting victim $t$) follows a multivariate mixture distribution consisting of $\mu_t$, $\mu_y$, and $e$. 

Under the trigger strategy of \ourattack, the discrepancy of the first-moment information might become less discriminative. As shown in Table~\ref{tab:FPRs and TPRs of defenses on MNIST, CIFAR-10, and GTSRB under SSDT attacks}, SCAn suffers on the MNIST and CIFAR-10 datasets with a much decreased TPR.

\noindent\textbf{Beatrix.} Beatrix \cite{ma2022beatrix} serves as a defensive mechanism against dynamic trigger backdoor attacks in neural networks. 
The idea is to utilize high-order statistics to capture the subtle difference between normal and backdoored representations in Euclidean space.
To this end, it resorts to the Gramian matrix, which represents both individual channel features and cross-channel correlations. 
Despite the fact that dynamic trigger attacks are hard to be detected by existing detection methods due to the {statistics of latent features} have been changed and the high confidence in the misclassification of $x \oplus g(x)$, the Gramian matrix has been found to be effective in discerning the nuance between clean and dynamic triggered samples under the vanilla dynamic trigger attack cases in \cite{ma2022beatrix}.

\begin{figure}[t!]
  \centering
  \includegraphics[width=0.3\textwidth]{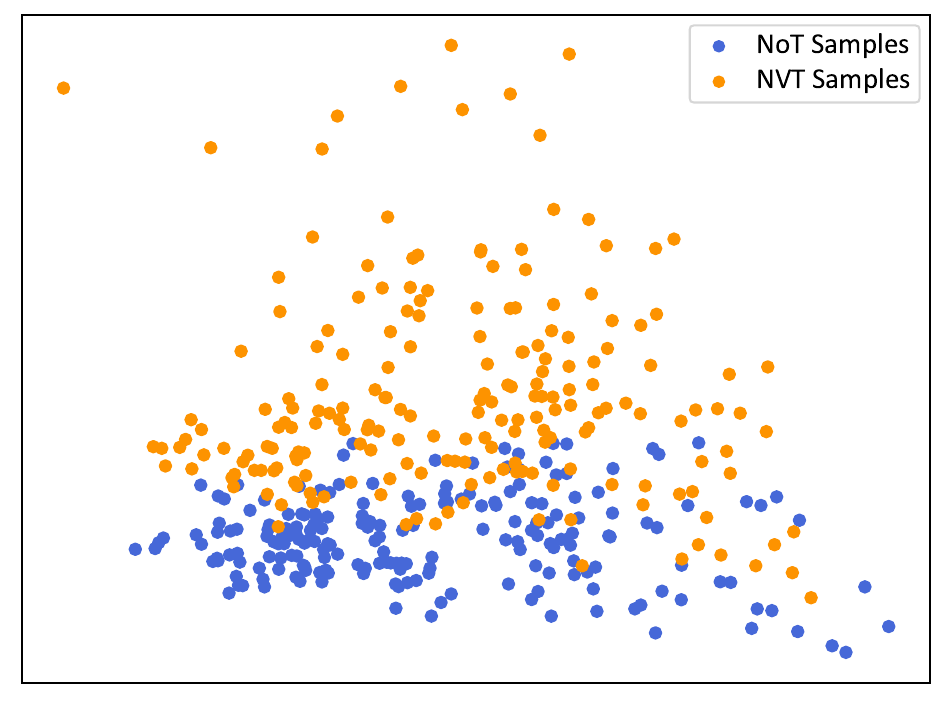}
 \caption{
 Top 2 principal components
 of Gramian features for NVT samples and NoT samples on GTSRB.}
  \label{fig:beatrix_gramian_pca}
  \vspace{-1mm}
\end{figure}

However, as depicted in Fig.~\ref{fig:beatrix_gramian_pca}, Beatrix faces challenges with \ourattack backdoor attacks. Under \ourattack, the Gramian feature representations of NVT samples and NoT samples remain distinctly separate in Euclidean space, yet neither set is capable of activating the backdoor, thus significantly impeding Beatrix's discriminative capabilities. 
Despite its notable robustness against dynamic backdoor attacks, Beatrix's effectiveness is substantially compromised  {when} source-specific strategy is further applied as in \ourattack.
As reported in Table \ref{precision}, even using high-order statistics,  channel-wise and cross-channel information to amplify the differences between normal samples and malicious samples in Euclidean space and reflect them on Gramian features, Beatrix still fails to yield satisfactory detection results.

%% file: sections/4_TED.tex
To overcome the limitations of existing defenses, we propose a novel backdoor detection approach that 
hinges on
the evolution dynamic of 
input samples from the front to the back of the neural network.
We formulate a new concept, 
the topological evolution dynamic (\ourdefense),
which differs from the vanilla Euclidean distance metrics in the existing approaches in two ways. 
First, instead of relying on a static sample representation, \ourdefense captures the input-to-output dynamic of a deep learning model.
Second, instead of relying on a fixed metric, \ourdefense leverages more robust (topological) neighbor relations among samples. 
Building on this foundation, we develop a detector that distinguishes potentially compromised samples from clean ones by comparing the ranking of the nearest neighbor from the predicted class that they follow throughout the layers of the network. 

In this section, we first provide an overview of the intuition, and key idea of \ourdefense, as well as the challenge for input-level detection. Then we further detail {the feature modeling of \ourdefense and how we build the detector based on the features obtained}.

\subsection{Intuition, Challenge, and  Key Idea}
\noindent\textbf{Intuitions on input-level backdoor detection.} 
A key intuition is that the network behaves differently on clean samples of the targeted label $t$ and the trigger-carrying samples targeted at that class label $t$, prompting us as humans to categorize the former as normal and the latter as abnormal. 
To this end, the backdoor detection problem is an anomaly detection task.
There are various forms of anomaly detection methods that have been utilized for backdoor detections, \ie, statistical methods~\cite{rousseeuw2018anomaly}, clustering methods~\cite{syarif2012unsupervised}, density-based methods~\cite{tang2017local}, \textit{etc}.
However, many of them formulate the difference from the perspective of feature representations, in which they believe the outputs from the penultimate layer $f_{N-1}$ of normal and abnormal inputs are separable.
Such an assumption may be compromised if encountering more advanced attacks, such as \ourattack.
Another intuition is that the clean samples of class $t$ and the trigger-carrying samples targeted at $t$ are significantly different in the input space, however, they are all classified as $t$ by the backdoored model in the end.
In a sense, they must evolve differently as they propagate through the network. 
This encourages us to model the evolution difference between normal and malicious samples for detection.



\noindent \textbf{Challenge on input-level detection.} We note that the previous backdoor detection methods largely focus on feature discrimination on certain layer(s), where the difference between normal samples and malicious samples are modeled in some metric space, particularly evaluated by the Euclidean distance of the penultimate layer. 
On the other hand, the way to implement anomaly detection for backdoor defense, in which the feature representations are further exploited, is in various forms.
Along with the advance of strong attacks, the effort to effectively discriminate malicious samples is becoming more extravagant.
For vanilla static-trigger attacks such as BadNet, the trivial clustering-based anomaly detection methods  \cite{tran2018spectral,chen2018detecting} can already achieve nearly 100\% detection accuracy and F1 score for every class. 
However, under stronger adversaries who utilize the source-specific design or the dynamic triggers, the individual feature representations of malicious and normal samples are inseparable, which violates the assumption of clustering-based methods. 
To this end,  first or higher-order statistics, 
which capture the discrepancy from the distribution of samples, are utilized in advanced methods like SCAn and Beatrix.
Nevertheless, as we illustrated in Sec. \ref{sec:limitations}, all these efforts cannot suffice to effectively defend \ourattack.

We conjecture that this is because these metric space-based methods merely examine the difference from latent (or raw) features, failing to capture the fundamental working discrepancy between malicious samples and normal samples throughout the network. 
The Euclidian nuance in SCAn and Beatrix in the penultimate layer is diminished under stronger adversaries.
Even employing advanced techniques like two-subgroup untangling and higher-order statistics in SCAn and Beatrix to amplify the difference, there still remains a significant challenge in developing a once-for-all solution.

\noindent\textbf{Key idea of topological evolution dynamics.} To overcome the challenges posed by existing solutions, we utilize the discrepancy evolution dynamics throughout the neural network, which examines input samples simultaneously from front to back in the network. 
We find that the topological evolution dynamics between these two groups of data points from the front to the end of the neural networks are significantly different in nature.
Our method considers not just the representation from the penultimate layer, but also the activations of multiple intermediate layers throughout the network. 
These activations are further exploited for topological analyses given activations of pre-stored benign samples. 
{Specifically, we consider the predicted class as the reference, and in each layer, we sort the database based on their activation distance to the input sample in that layer. We then record the ranking of the nearest neighbor from the reference class in each layer. The list of rankings from all considered layers forms the feature for discrimination, reflecting the evolution dynamics of the input sample.}

Intuitively, benign samples should exhibit more consistent rankings than malicious samples in the list since the predicted class is legitimate for benign samples,  so samples from the same class should be nearer neighbors for them in each layer.
In contrast, the predicted class for malicious samples should give a wrong reference for earlier layers in the network,  and the reference becomes legitimate only in the end.
The rationale for this approach is two-fold. First, the activations at the end of a backdoored network for malicious samples $x \oplus \delta$ should be associated with the target label $t$, making them closer to the benign samples from $X_t$ in terms of the semantic features. Second, $x \oplus \delta$ is somewhat identical to $x \in X_s$ in terms of their shallow features in earlier layers, such as shape and texture, making them appear closer to the benign samples of the original class of $x$. 
{It should be noted that this also applies to the cases in advanced attacks like source-specific backdoor attacks. 
NVT samples should have more consistent rankings than VT samples successfully associated with the target label since triggers do not change the shape and texture of samples for inconspicuousness, yet VT samples fall into the ``wrong'' predicted class in the end.}
By leveraging these insights, it turns out that simple outlier-detecting techniques such as a PCA-based detector 
 can suffice to effectively discriminate between benign and malicious samples since their evolution dynamics are significantly different already.

\begin{figure*}[t]
	\centering
	\subfloat[MNIST]{\includegraphics[width=.21\linewidth]{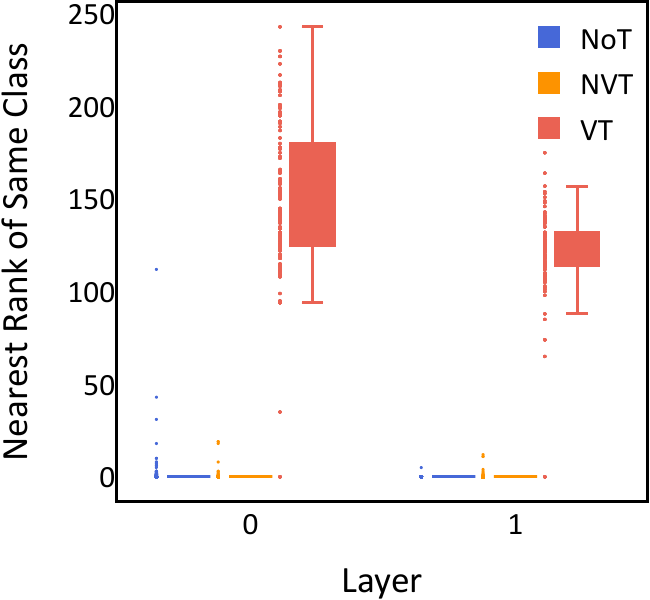}}
    \hspace{4pt}
    \subfloat[CIFAR-10]{\includegraphics[width=.765\linewidth]{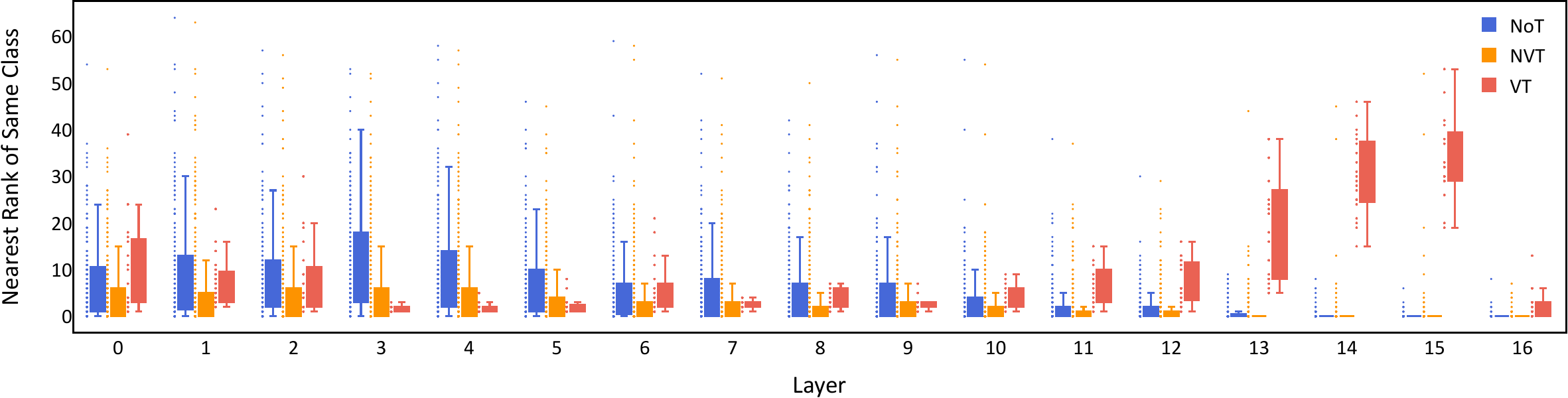}}\\
	\subfloat[GTSRB]{\includegraphics[width=.74\linewidth]{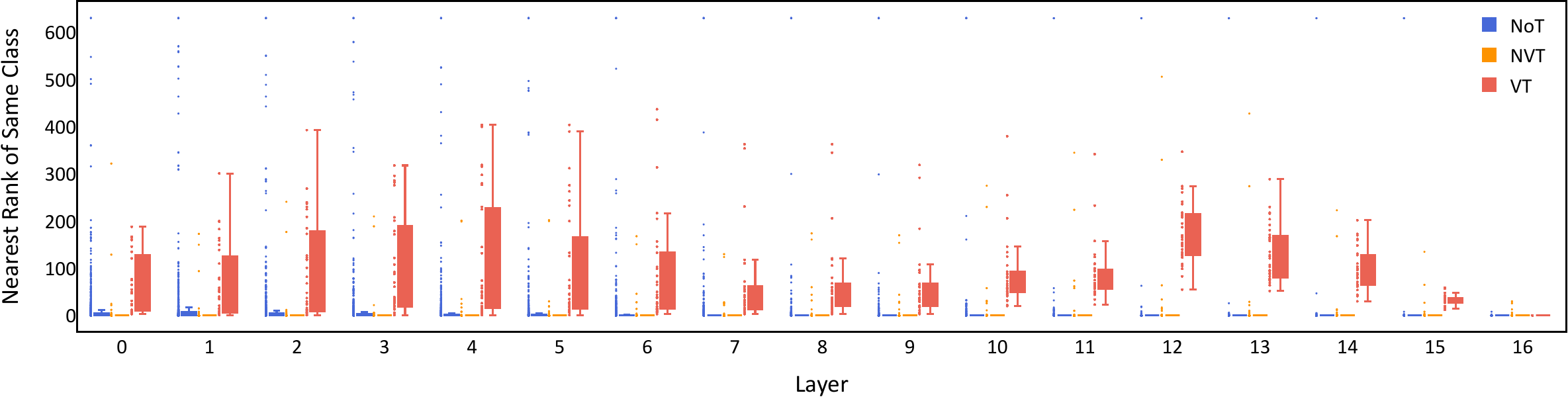}}
	\caption{Box plots of the topological feature vectors on the three datasets.} 
 \label{fig:TopologyPersistenceDiagram}
 \vspace{-1mm}
\end{figure*}

\subsection{Feature Modeling via Topological Evolution}
\label{subsec:topologicalFeature}
\textit{Topological space induced by metric space.} Formally, a  vector (or matrix) set $\mathcal{V}$ and a metric function $d$ can form a metric space $(\mathcal{V}, d)$, where $d$ is defined on $\mathcal{V}$: $\mathcal{V} \times \mathcal{V} \rightarrow [0, \infty )$. 
Previous input-level detection methods all follow this modeling. 
Particularly,  denote the representation output of an input sample $x$ at the $l$-th layer of the deep learning model as  $h_l(x) = v \in \mathcal{V}^{(l)}$, the outputs of the $l$-th layer for all samples make up the set under consideration $\mathcal{V}$. 
Typically, previous detection methods utilize the penultimate layer or the last layer to form a single metric space $(\mathcal{V}, d)$, and  $d$ is usually the Euclidean distance function.
As we have demonstrated in Sec.~\ref{Sec:SSIDattack}, merely modeling latent features from one layer and examining them solely based on Euclidean distance is insufficient. There may always be more advanced attack strategies, like or better than \ourattack, that can blend the representations of malicious attack samples and normal samples better.

Therefore, alternatively, we approach the modeling from a topological perspective. 
Instead of relying solely on the distance between feature vectors, we focus on capturing the relative closeness of feature instances to determine how natural a sample is to its predicted class.
Specifically, we can induce a topological space based on the standard metric distance. 
Given any $v \in \mathcal{V}$ and  the distance function $d$,  if we pre-set a radius $r$, we can define an open ball  around $v$, which is  $\mathcal{B}(v,  r)= \{v' \in \mathcal{V}| d(v, v') < r \} \subset \mathcal{V}$. 
Thus, we can obtain a topology that is the collection of all the open balls. 
Note that, each open ball $\mathcal{B}(v,  r)$ defines a neighborhood around $v$ that consists of all the points that are ``close" to $v$, where $r$ reflects how close they are.

\textit{Neighborhood in each layer.} As mentioned earlier, we believe that benign input samples should be close to some other benign samples from the predicted class, whether model $f$ is backdoored or not. 
Thus, for a benign sample (not an outlier) $x_u$ with predicted label $y_u$ and its latent feature $v_u^{(l)} \in \mathcal{V}^{(l)}$ at each layer $f_l$, there exists another benign sample $x'$ with the same label that is close to it.
In other words, we have an intuition  as follows:
$$\forall \ r_l \geq  r_l^*, \exists \ x'  \in X_{y_u} -\{x_u\} \text{ satisfying }  h_l(x') \in \mathcal{B}(v_u^{(l)}, r_l), $$
where $r_l$ denotes the radius at $l$-th layer, and its lower bound $r_l^*$ is relatively small.
{Particularly, if $h_l (\cdot)$ is continuous everywhere in a  continuous input space $X$, we can have $r^*_l \approx 0$ since the distance $\|x' - x_u\|$ could be arbitrarily close to 0 for some $x'$.}
Given the non-continuous input space like the image space, the lower bound $r^*_l$ by definition is controlled by the nearest neighbor from $X_{y_u} - \{x_u\}$.

\textit{{Evolution dynamic of the neighbors}.}
Intuitively, a benign sample (not an outlier) should be correctly associated with its predicted class, \ie, its neighborhood should be full of the samples from the predicted class. 
Moreover, in general, benign samples from the same class tend to remain clustered together rather than becoming separated, from the input layer towards the output layer. 
Research has shown that deep neural networks tend to learn a maximally compressed mapping of the input that preserves as much information about the output as possible~\cite{7133169}, and the network evolves by pulling similar samples into tighter clusters while separating different classes~\cite{papyan2020prevalence}. 
This indicates that, on average, a benign sample consistently becomes `nearer' to its neighbors belonging to the same predicted class throughout the layers of the network. 
Further, we conjecture that a malicious sample, even if it ultimately falls into the target class, should appear bumpy on its way toward the end, and may activate neurons from its original class (its open ball comprised representation of samples from the original class samples) in the intermediate layers.

\textit{Modeling the dynamic through ranking.} 
{Based on the above analysis, we model how the closeness of the input sample $x$ to the predicted class evolves as it passes through the network. We then use this dynamic to distinguish between the malicious and benign samples.
Particularly, to represent the closeness of the input sample $x$ to the predicted class, we utilize $x$'s nearest neighbor from the predicted class at each layer.
The rationale is if the closeness to the nearest neighbor at each layer is diminished, the sample $x$ is certainly shifting away from the predicted class.
To quantify the closeness to the nearest neighbor, one straightforward solution is to measure the distance between them, \ie, $r_l^*$.
}
However, the quantity of these distances from different layers measured in the metric space themselves might not be comparable since the dimension of each layer's output is variant, \ie, they lie in different metric spaces of different dimensions.
Therefore, we cannot model this dynamic simply through the distance metric. Alternatively, we use the ranking of the nearest neighbor from the predicted class at every layer considered to model this dynamic.

Formally, for a $c$-classes classifier, we  randomly select  $m$ samples from each class, resulting in $c\times m$ samples in total. 
For each input sample $x$, we can obtain a ranking list \textit{w.r.t.} these $c\times m$ samples, sorted by their representation distances to $x$ at each layer.
We record the rank of $x$'s nearest neighbor from the predicted class at layer $l$ as $K_l$.
As a result, we can obtain a sequence of ranks that represent how close {$x$} is to its nearest neighbor from the predicted class at each layer: [$K_1$, $K_2$, ... $K_l$ ..., $K_N$].
This sequence describes the topological evolution dynamic of the input sample $x$ \textit{w.r.t.} the predicted class as it tracks the minimal closeness of $x$ towards the predicted class throughout the network. 
{We build an outlier detector upon the ranking sequences of normal samples using a PCA model. 
The PCA model takes the ranking sequences from all $c\times m$ samples and a reject parameter $\alpha$ as inputs. Then it will compute a threshold $\tau$ that preserves the $1 - \alpha$ percentage of the principal components \textit{w.r.t.} these samples, and treat the remaining $\alpha$ percentage samples as outliers. 
Finally, the threshold $\tau$ is  used for detection.
In practice, we use the open-source code of the Python Outlier Detection (PyOD) library\footnote{PyOD library: \url{https://github.com/yzhao062/pyod}}.
}
Fig.~\ref{fig:TopologyPersistenceDiagram} box-plots these feature vectors for different types of samples on the three datasets. Algorithm~\ref{algo:ted} illustrates the complete process of \ourdefense.

\begin{algorithm}
Given a $c$-class neural network model $f$ with $N$ considered layers, $m$ samples from each class, a metric distance function $d$, a PCA model $\mathtt{PCA}( \cdot , \alpha)$ with {reject} parameter  $\alpha$, a sample set $X_{test}$ to be detected.

\tcc{Store latent features}
\textbf{{Initialize}} $S_1=S_2=...=S_c=\emptyset$

\For{$i=1$ to $c$}{
    \For{$j=1$ to $m$}{
        $x \leftarrow \mathtt{random\_sample}(X_i)$\\
        Stack $x$ in $S_i$;\\
        $[h_l(x )]_{l=1}^N  \leftarrow$ Forward  $x$ to  $f$
    }
}

\tcc{Store rank lists}
\For{$i=1$ to $\|S\|$}{
    $j = \argmax_{k \in [1,c]} f(x_i)_k$
    
    \For{$l=1$ to $N$}{
    $S_\mathtt{sorted} = \mathtt{sort\_by\_distance}(d, h_l(\cdot), S, x_i)$
    
    $x_\mathtt{nn}= \mathtt{get\_nearest\_neighbor}(d, h_l(\cdot), S_j-x_i, x_i)$
    
    $K_l^{(i)} = \mathtt{get\_rank}(S_\mathtt{sorted}, x_\mathtt{nn})$
    }
    Record $[K_l^{(i)}]_{l=1}^N$
}

\tcc{Build detector with rank lists}
\textbf{Fit} PCA model $\mathbf{M} = \mathtt{PCA}(\{[K_l^{(i)}]_{l=1}^N\}_{i=1}^{\|S\|} , \alpha)$

$\tau = \mathbf{M}.\mathtt{get\_detect\_threshold}(\alpha)$

\tcc{Detect sample with threshold}
\textbf{{Initialize}} malicious samples set  $X_{malicious}=\emptyset$

\For{$x$ in $X_{test}$}{
\If{ $\mathbf{M}(x) > \tau$}{
Add $x$ in $X_{malicious}$
}
}
\Return $X_{malicious}$

\caption{The overall algorithm of \ourdefense}\label{algo:ted}
\vspace{-1mm}
\end{algorithm}

%% file: sections/5_Experiment_1.tex
This section details the comprehensive suite of experiments conducted to evaluate the robustness of our proposed \ourdefense methodology against a spectrum of settings.
We first scrutinize the performance of \ourdefense against various forms of attacks in Secs.~\ref{sec:evaluation} and \ref{sec:diverse-attack-evaluation}, including \ourattack, dynamic-input backdoor attack~\cite{nguyen2020input}, and two types of source-specific attacks ---TaCT~\cite{tang2021demon} and composite backdoor attack~\cite{lin2020composite} by comparing \ourdefense with a collection of state-of-the-art defense mechanisms (SentiNet~\cite{chou2018sentinet},  STRIP~\cite{gao2019strip}, SCAn~\cite{tang2021demon} and Beatrix~\cite{ma2022beatrix}). For \ourattack, the datasets and models presented in Sec.~\ref{subsec:DataModel} are used. {For other attacks, we make use of the datasets and models presented in their original paper.} 
In Sec.~\ref{sec:adaptive-attack}, we study the resistance of \ourdefense against adaptive attacks. By default, we take all the outputs of the Conv2D layers to extract topological feature vectors.  
In the ablation study (Sec.~\ref{subsec:AblationStudy}), we further investigate the influence of layer types and quantities on \ourdefense.  
{In Sec.~\ref{subsec:ComplexDataset}, we corroborate the efficacy of \ourdefense on complex datasets (\eg, ImageNet) and different applications (\eg, NLP).}
Networks without Conv2D layers and scenarios involving models pending backdoor detection are discussed in {Appendix-C} and {Appendix-D}, respectively. 

\begin{figure}[t]
  \centering
  \includegraphics[width=0.45\textwidth]{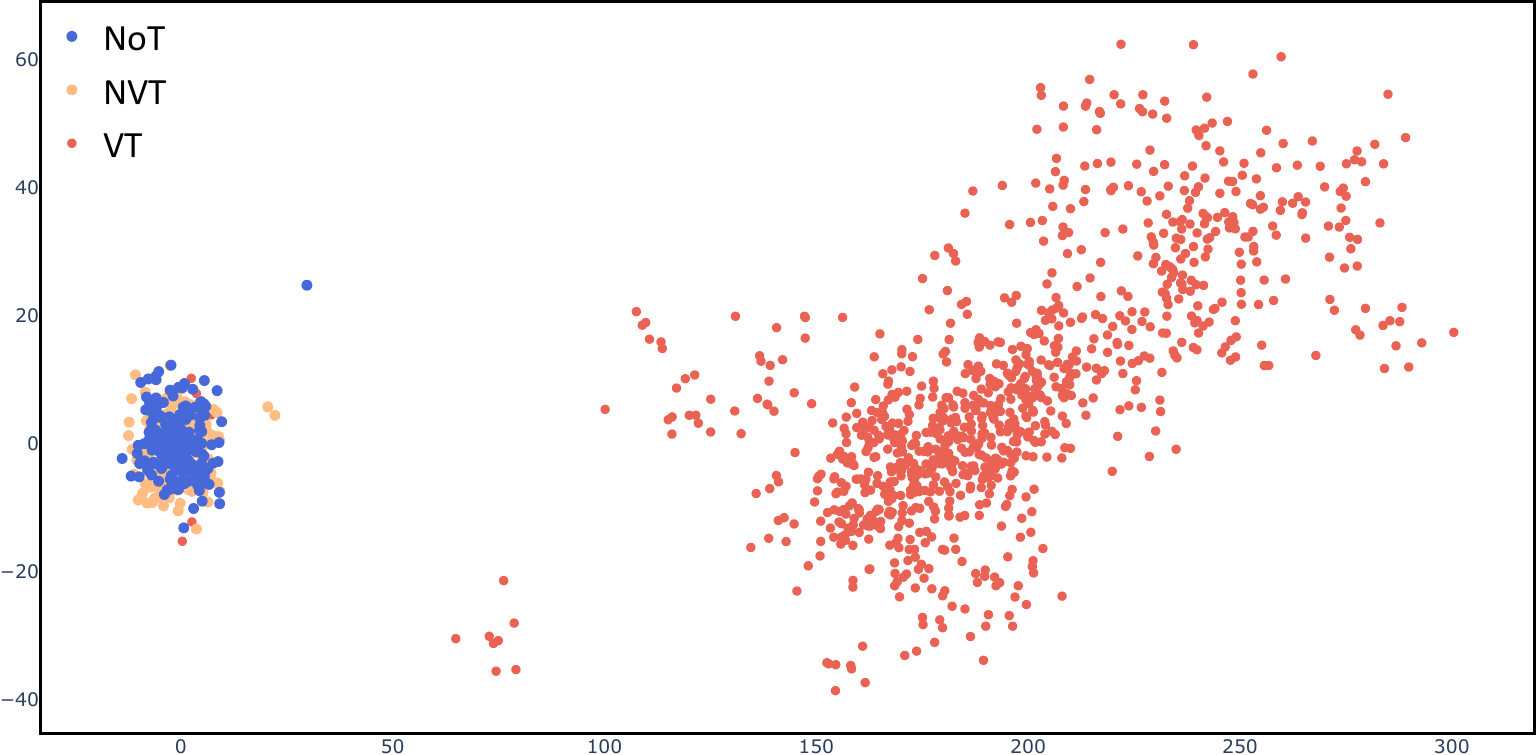}
 \caption{PCA plot of topological features from various types of MNIST samples.}
  \label{fig:TED_Flowchart}
\end{figure}

\subsection{Effectiveness of \ourdefense Against \ourattack}
\label{sec:evaluation}
As listed in Table~\ref{tab:FPRs and TPRs of defenses on MNIST, CIFAR-10, and GTSRB under SSDT attacks}, \ourdefense achieves 100\% TPR rate in detecting VT samples of \ourattack for all the three datasets. To better visualize the reason why \ourdefense performs superior in detecting \ourattack attacks, we make a PCA plot of the extracted features similar to Fig.~\ref{fig:beatrix_gramian_pca} {and present the result in Fig.~\ref{fig:TED_Flowchart}}. It is clear from this figure that features of various types of samples (\ie, VT v.s. NVT/NoT) extracted from the topological structure are clearly separable. This substantiates the suggested topological features can effectively capture the subtle differences between clean and malicious samples.

\begin{table}[h]
\centering
\caption{\label{tab:Accuracy comparison of defense methods} Accuracy (\%) of SOTA detectors against \ourattack.}
\begin{tabular}{@{}llllll@{}}
\toprule
Dataset   & \textbf{\ourdefense} & \textbf{Beatrix}  & \textbf{SCAn}     & \textbf{STRIP}    & \textbf{SentiNet}  \\ \midrule
MNIST     & 97.99   & 89.05  & 69.50  & 47.88  & 49.13   \\
CIFAR-10  & 97.63   & 82.30  & 65.75  & 46.75  & 51.00   \\
GTSRB     & 98.63   & 83.34  & 97.25  & 48.63  & 50.00   \\ \bottomrule
\end{tabular}
\end{table}

\begin{table}[h]
\centering
\caption{\label{tab:Precision comparison of defense methods} Precision (\%) of SOTA detectors against \ourattack.}
\begin{tabular}{@{}llllll@{}}
\toprule
Dataset   & \textbf{\ourdefense} & {\textbf{Beatrix}}  & \textbf{SCAn}     & \textbf{STRIP}    & \textbf{SentiNet}  \\ \midrule
MNIST     & 96.90   & 94.93  & 89.80  & 9.52   & 37.93   \\
CIFAR-10  & 95.47   & 77.99  & 87.95  & 0        & 58.70   \\
GTSRB     & 98.63   & 75.00  & 95.22  & 0        & 50.00   \\ \bottomrule
\end{tabular}
\label{precision}
\end{table}

We further measure the accuracy and precision of \ourdefense and other SOTA detectors and report the results in Tables~\ref{tab:Accuracy comparison of defense methods} and~\ref{tab:Precision comparison of defense methods}. From these tables, both STRIP and SentiNet exhibit about 50\% detection accuracy (\ie, random guess), failing to detect  \ourattack backdoor totally. Moreover, SCAn is inferior to Beatrix and \ourdefense, since it is designed for resisting source-specific but trigger-static backdoors, and \ourattack violates its trigger strategy. Beatrix suffers from the issue of high false positives (low precision as shown in Table~\ref{tab:Precision comparison of defense methods}) under \ourattack, making its detection accuracy 13\% lower than \ourdefense on average.

\subsection{Evaluating \ourdefense Against Diverse Attacks}
\label{sec:diverse-attack-evaluation}
Noted that SCAn is customized to detect a  source-specific backdoor called TaCT (targeted contamination attack) \cite{tang2021demon}, and Beatrix is tailored to defeat a trigger-dynamic backdoor \cite{nguyen2020input}. We assess \ourdefense against these attacks and their variants. {The results are reported on CIFAR-10 and GTSRB as customized detectors demonstrate 100\% TRP and 0\% FPR on MNIST. For the same reason, we do not include the result of source-agnostic or static-trigger attacks \cite{gu2017badnets,chen2017targeted,liu2017trojaning,zeng2021rethinking}.}

\textit{Dynamic-input backdoor attack \cite{nguyen2020input}.}
\label{Dynamic-input backdoor attack}
This attack works under the same threat model as \ourattack. It co-optimizes an encoder-decoder network $g$ together with the to-be-backdoored model $f$ to generate dynamic trigger-carrying samples (Eq.~\ref{Eq:DynamicTrigger}). Beatrix enlarges the subtle difference between normal and dynamically-triggered samples in Euclidean space by using the Gramian matrix to record their latent features’ correlation and their high-order statistics.

We use the code provided by \cite{nguyen2020input} and report the detection results in Table~\ref{tab:Conventional Dynamic Input Backdoor Attack Comparison}. From the results, it is clear that \ourdefense is comparable to Beatrix in resisting traditional trigger-dynamic backdoor attacks. {As shown in Fig.~\ref{fig:TopologyPersistenceDiagram_DynamicInput} of {Appendix-A}, the box plot of the topological feature vectors under this attack \cite{nguyen2020input} also supports this conclusion.

\begin{table}[h]
\centering
\caption{\label{tab:Conventional Dynamic Input Backdoor Attack Comparison} Detection performance against the attack in~\cite{nguyen2020input}.}
\begin{tabular}{lcccc}
\toprule
& \multicolumn{4}{c}{Dataset} 
\\
\cmidrule(lr){2-5}
 & \multicolumn{2}{c}{CIFAR-10} & \multicolumn{2}{c}{GTSRB} \\\cmidrule(lr){2-3} \cmidrule(lr){4-5}
 & \textbf{\ourdefense} & \textbf{Beatrix} & \textbf{\ourdefense} & \textbf{Beatrix} \\
\midrule
{\textbf{TPR (\%)}} & 91.60 & 99.00 & 98.00 & 99.80 \\
\textbf{Precision (\%)} & 99.34 & 95.40 & 100.00 & 99.80 \\
\textbf{F1 (\%) }& 95.31 & 97.20 & 98.90 & 99.80 \\
\bottomrule
\end{tabular}
\end{table}

\textit{Targeted contamination attack~\cite{tang2021demon}.}
\label{Targeted contamination attack} 
TaCT is the first to explicitly use the source-specific trigger strategy to disperse the predominant effect of the trigger, while such an effect is the key enabler for detectors like Neural Cleanse \cite{wang2019neural}, SentiNet \cite{chou2018sentinet} and STRIP \cite{gao2019strip}. 
In particular, TaCT is implemented by stamping the same static trigger (\eg, a small square) on the samples from the victim source class to prepare the backdoor dataset $D_b$ (as in Eq.~\ref{Eq:backdoor dataset}) and on the samples from the non-victim class to prepare the laundry dataset $D_l$ (as in Eq.~\ref{Eq:laundry dataset}). 
And SCAn is customized to address this challenge by observing that, after conducting decomposition on benign and malicious samples, their features are separable in terms of the first moment.

We follow the same settings used in TaCT to launch this attack, and the detection performance of SCAn and \ourdefense are listed in Table~\ref{tab:Conventional Source-Specific Backdoor Attack Comparison}.
{From this table, it is evident that \ourdefense demonstrates comparable performance to SCAn on TPR and FPR.
We further note that as mentioned in \cite{tang2021demon,ma2022beatrix}, SCAn requires to discern about 50 VT samples before it can reliably detect further samples, meaning that in the online setting, SCan would likely miss the first dozens of VT instances. \ourdefense does not suffer from this cold-start problem. 
}
{Fig.~\ref{fig:TopologyPersistenceDiagram_TaCT} in Appendix-A further depicts the topological feature vectors under TaCT}.

\begin{table}[h]
\centering
\caption{\label{tab:Conventional Source-Specific Backdoor Attack Comparison}
Detection False Positive Rate (\%) against TaCT.}
\begin{tabular}{lllll}
\toprule
&
\multicolumn{4}{c}{Dataset}
\\
\cmidrule(lr){2-5}
& \multicolumn{2}{c}{CIFAR-10}    & \multicolumn{2}{c}{GTSRB}               \\ \cmidrule(lr){2-3} \cmidrule(lr){4-5} 
 & {\textbf{\ourdefense}}& \textbf{SCAn} & \textbf{\ourdefense} & \textbf{SCAn}               \\ \midrule
95\% TPR                     & 0.75            & 0.47          & 0            & 0.32 \\
99\% TPR                     & 2.00         & 0.48          & 0.90         & 1.10 \\ \bottomrule
\end{tabular}
\end{table}

\textit{Composite backdoor attack~\cite{lin2020composite}.}\label{Composite backdoor attack} It is a variant of the source-specific attack. It differs from TaCT in how the backdoor dataset $D_b$ (Eq.~\ref{Eq:backdoor dataset}) and the laundry dataset $D_l$ (Eq.~\ref{Eq:laundry dataset}) are prepared. When preparing $D_b$, a small area (\ie, a cutout) of the sample from $X_s$ is stamped on itself; when preparing $D_l$, a cutout of the sample from non-victim classes is also stamped on itself (the \textit{mixer} function in \cite{lin2020composite}). That is said, the trigger patterns for $D_b$ and $D_l$ are different. 

We follow the exact settings employed in \cite{lin2020composite} and conduct experiments on CIFAR-10. This attack achieves 85.62\% accuracy on benign samples and 84.25\% attack success rate. \ourdefense achieves TPR of 94.73\% at FPR of 5\%, along with an AUC score of 0.9849, compared to Beatrix with 92.11\% TPR at 5\% FPR and 0.9688 AUC score.

\begin{figure}[ht]
\includegraphics[width=0.46\textwidth]{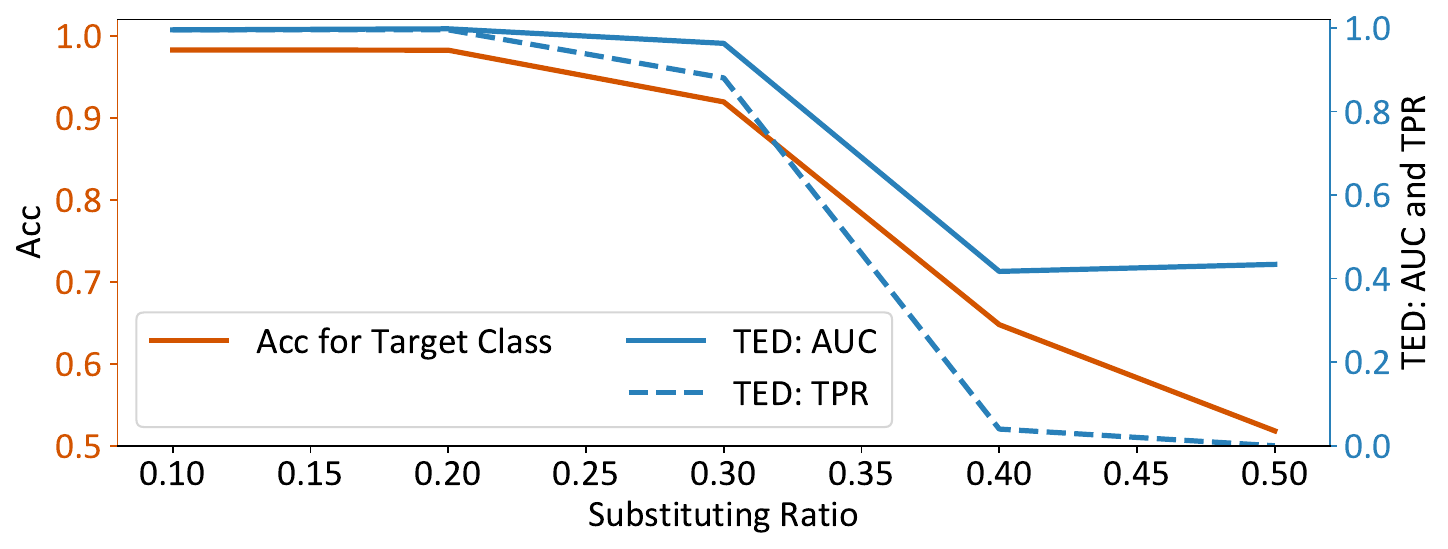}
\caption{\label{fig:adaptive_attack}{Accuracy, AUC, and TPR v.s. substituting ratio}.}
\end{figure}

\subsection{Adaptive Attacks on \ourdefense}
\label{sec:adaptive-attack}
A crucial aspect of evaluating the robustness of a backdoor detector is its resistance to adaptive attackers, who are aware of the defense mechanism and aim to bypass detection. Indeed, under the trigger-dynamic strategy (\textit{but not source-specific}), Beatrix \cite{ma2022beatrix} investigates adaptive attackers who try to minimize the difference in  Gramian features between benign samples and malicious samples when embedding a backdoor. Similarly, under the source-specific 
 strategy (\textit{but not trigger-dynamic}), SCAn \cite{tang2021demon} studies how to optimize a better trigger that can bypass detection. From this sense, the designed \ourattack attack is just an adaptive attack that breaks the assumption of the latent separability in the metric space, used by \cite{tang2021demon,ma2022beatrix} and many more. So, the question is: \textbf{will \ourdefense make a real difference when resisting adaptive attacks?} 

 To answer that question, we first follow the same wisdom from earlier works to design adaptive attacks. That is, the adaptive attacker aims to optimize for a better attack by designing an appropriate loss function other than  the original backdoor learning loss $\mathcal{L}_o$. Given the non-invertible and non-differentiable nature of our topological feature extraction method, the attacker cannot directly optimize the topological features for malicious samples as before. Alternatively, the attacker studies loss functions that can allow malicious samples to mimic the topological features of clean samples. 

In this regard, we formulate the following three heuristic loss designs:
\begin{IEEEeqnarray}{rCl} 
\mathcal{L}_{\text{total}} &= & \mathcal{L}_o + \lambda_1 \mathcal{L}_1, \nonumber \\ 
\mathcal{L}_{\text{total}} &= &\mathcal{L}_o + \lambda_2 \mathcal{L}_2, \nonumber \\
\mathcal{L}_{\text{total}} &= &\mathcal{L}_o + \lambda_3 \mathcal{L}_3. \nonumber 
\end{IEEEeqnarray} 
Here, $\mathcal{L}_1$ is inspired by the distance metric learning for large margin nearest neighbor classification \cite{weinberger2009distance}, which is designed to minimize the Euclidean distance between poisoned activations within the same class while maximizing it between different classes, thus mimicking the topological feature of a benign sample. $\mathcal{L}_2$ is inspired by the K-means algorithm \cite{MacQueen1967SomeMF}, which is designed to reduce the distance from a given point (the poisoned activation) to the geometric centroid of the target class. {And $\mathcal{L}_3$ is inspired by the latest advancement of NLP backdoor \cite{li2021backdoor}. In this case, the trigger-carrying samples are forced to match the clean samples from the target class only at the first few layers, with the aim of making topological features associated with deeper layers appear normal.} The formal expressions of these three loss functions are given {in Appendix-B}.

We implement the above three adaptive attacks on {MNIST} with a poison rate 2\% and $\lambda_i =1~(i \in [1, 3])$ using {ResNet-18\footnote{We exclude the usage of shallow models since the rationale of $\mathcal{L}_3$ is incompatible with shallow models.}}. 
After successfully reducing the loss $\mathcal{L}_{\text{total}}$ after 10,000 iterations, {the model's accuracy and attack success rate remain high, sustaining levels above 99\%}. 
Surprisingly, \ourdefense demonstrates remarkable resilience against these three adaptive attacks, achieving an AUC of 0.99 and a TPR of 100\%. 
We attribute this to that neither of the three loss functions $\mathcal{L}_1$, $\mathcal{L}_2$ and $\mathcal{L}_3$ can truly break the feature separability in the topological space since it captures the very nature of benign and malicious samples. As an example, we empirically observe, after applying $\mathcal{L}_3$, the topological features exhibit distinctive dynamics in the middle layers. 


Given the complexities associated with manipulating the loss function, we turn to the strategy of manipulating the training data. {In this approach, a certain percentage (\ie, the substitution ratio) of the clean samples from the victim class is re-labeled as the target}. 
{This equals replacing the training data with specific dirty/noisy labels, which has been extensively studied in the ML community~\cite{song2022learning}.} 
The rationale of this adaptive attack is that samples with dirty labels will gather around the trigger-carrying samples, creating an innocent-looking topological feature as it propagates.

We conduct the experiment using the same setting as above,  
and the results are depicted in Fig.~\ref{fig:adaptive_attack}.
Despite the substitution, the model maintains a stable attack success rate of 99.5\%, even as the substitution ratio increased. 
The rise in the substitution ratio directly correlates with a decline in the accuracy of benign samples from the target class, caused by the dirty labelling mechanism. In the worst case, when 50\% of the samples from the target class are mislabeled, the TPR of \ourdefense is only 50\% (random guessing). But the model accuracy for this target class is  only 50\%, which is also useless. Another observation is that \ourdefense maintains 100\% TPR until the substitution ratio exceeds 20\%. We argue this does not pose a real threat to \ourdefense since deliberately dirty-labeling over 20\% of one class in the training data can be easily spotted by many off-the-shelf methods \cite{song2022learning}.

\subsection{Ablation Studies of \ourdefense}
\label{subsec:AblationStudy}

At the heart of \ourdefense implementation is the 
layer-wise embedding, from which we extract the topological feature vector. As mentioned at the beginning of Sec.~\ref{Sec:exp}, we take all embeddings output by the Conv2D layers by default. Here, we further conduct a series of ablation studies to explore the properties of \ourdefense.  All the experiments are implemented under CIFAR-10 and GTSRB, with neural architecture varying from shallow models (\eg, 4-layer CNN) to deeper ones (\eg, PreAct-ResNet18), attacked by \ourattack by default. 

\begin{figure}[t]
    \centering
    \subfloat{\includegraphics[width=0.45\textwidth]{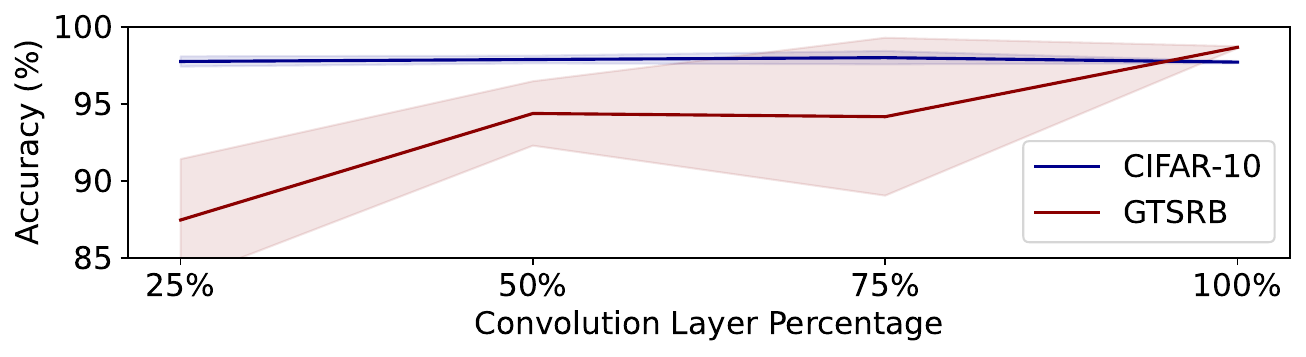}}
    \\
    \subfloat{\includegraphics[width=0.45\textwidth]{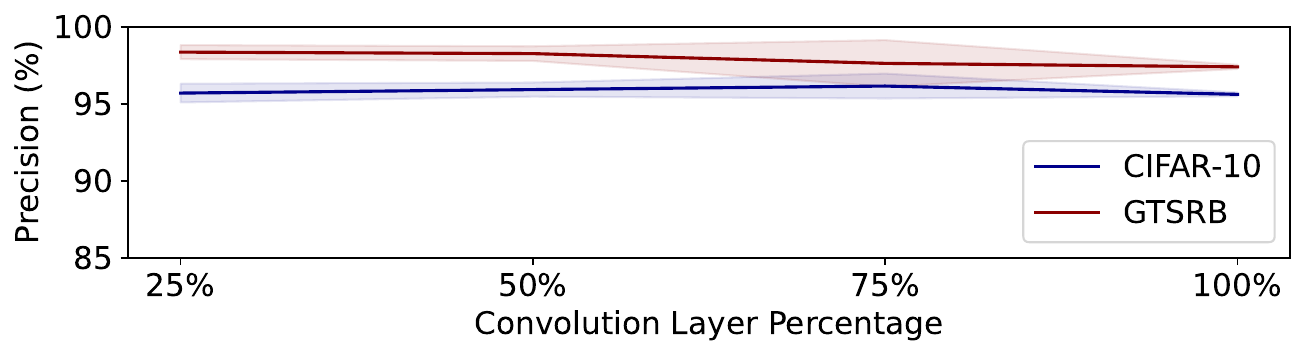}}
    \caption{Accuracy (top) and precision (bottom) with various percentages of convolution layers in \ourdefense.}
   \label{fig:Accuracy and Precision comparison in different percentage of TED snapshot}
\end{figure}
  

\textit{Layer-reduced variant of \ourdefense.} Though it is clear that \ourdefense will be more efficient than other online detectors like Beatrix \cite{ma2022beatrix}, SCAn \cite{tang2021demon}, and STRIP \cite{gao2019strip} due to its simplicity in feature extraction, it is always beneficial to squeeze efficiency in real applications. {For this purpose, we conduct a test by randomly sampling some Conv2D layers from PreAct-ResNet18 (measured in percentage), and depict the accuracy and precision of the layer-reduced version \ourdefense in  Fig.~\ref{fig:Accuracy and Precision comparison in different percentage of TED snapshot}.} 
It is observed that different layer-reduced versions maintain high accuracy (mostly over 90\%) and precision (over 95\%), and it might be possible that the efficiency of \ourdefense can be further improved.


\begin{table}[t!]
\centering
\caption{AUC of \ourdefense on CIFAR-10 with various layer combinations.}
\label{tab:TED_Performance_different_layer_combinations}
\begin{adjustbox}{width=0.49\textwidth}
\begin{tabular}{@{}lllll@{}}
\toprule
\textbf{Attack}    & \textbf{Model}              & \textbf{C} & \textbf{C \& R} & \textbf{C \& R \& L} \\ \midrule
\ourattack & PreAct-ResNet18              & .9953         & .9953                 & .9954                           \\
Composite \cite{lin2020composite} & PreAct-ResNet18              & .9494         & .9766                 & .9852                           \\
Composite \cite{lin2020composite} & 4-layer CNN & .7422         & .9466                 & .9849                           \\ \bottomrule
\end{tabular}
\end{adjustbox}
\flushleft Notes: C for Conv2D, R for ReLU, and L for Linear.
\end{table}

\textit{Effect of layers other than Conv2D.} We also add the ReLU and Linear layers for building the topological features, and the detection AUC of \ourdefense is reported in Table~\ref{tab:TED_Performance_different_layer_combinations}. For deep networks like PreAct-ResNet18, the detection AUCs are similar to each other under different settings, regardless of the attack method (\ie, \ourattack or Composite backdoor \cite{lin2020composite}). For shallow networks like 4-layer CNN, adding more layers does boost the detection performance. {We attribute this to the fact that the Conv2D operator in shallow networks is not trained to have strong feature extraction capability~\cite{mhaskar2017and}, making the topological features from these embedding spaces less discriminative.} As such, we should use the full layers for building \ourdefense if the network is shallow. We delay the discussion of detecting \ourattack backdoor for network without Conv2D layers to Appendix-C. 

\subsection{Further Evaluation on Other Tasks, Datasets, Models, and Attacks}
\label{subsec:ComplexDataset}

We further investigate \ourdefense's performance on more complex datasets and tasks, including advanced backdoor attacks in NLP tasks~\cite{yang2021careful}.



First, we test \ourdefense on the PubFig dataset and a subset of the ImageNet dataset with VGG16 as the target model \cite{simonyan2015deep}. 
The PubFig dataset presents a challenging scenario with 83 classes of 64×64 cropped facial images, emulating a facial recognition environment \cite{kumar2009attribute}. 
The subset of ImageNet includes 100 randomly selected classes, each with 500 images, following the experimental settings in \cite{ma2022beatrix, li2021invisible}. 
On VT samples, \ourattack can achieve 92.5\% accuracy in PubFig and 97.5\% in ImageNet, with over 99\% accuracy on NoT samples in both datasets. 

We extract topological features from the output of the Conv2D, ReLU, and Linear layers, using 200 inputs per label as the training set. \ourdefense shows promising results with a 0.9296 AUC on PubFig and 0.9972 on ImageNet. Additionally, in terms of the exploration of generalizability, when \ourdefense is trained on a limited subset with 200 images from 10 ImageNet classes, it still retains efficacy when tested on 10 entirely different classes, achieving an AUC of 0.9816, TPR of 93.5\%, and FPR of 3.5\%. 
Similarly, we box-plot the topological feature vectors for the NoT, NVT, and VT samples on these two datasets, and the result is shown in Fig. \ref{fig:TopologyPersistenceDiagram_pubfig_imagenet} of {Appendix-A}. 




Next, we go beyond the easier cases of feedforward networks with a clear concept of layers, and evaluate \ourdefense 
on more complex model architectures.
%
In particular, we assess the performance of \ourdefense against backdoor attacks on BERT-based models in NLP applications.

Due to the architectural difference of the transformer, we register forward hooks to record activations at the dense layer, self-attention layer, and word embedding layer. 
{And then we extract the topological feature vectors as usual (see Sec.~\ref{subsec:topologicalFeature}) with one exception: instead of recording the ranking of the nearest neighbor from the predicted class in each hook, we record the rankings of the $k$-nearest neighbors. We define the normalized version of $k$ as the radius $r =  \frac{k}{c\cdot m} * 100\%$. As before, $c$ is the number of classes in the classification task, and $m$ is the number of benign samples in each class.} 

We then evaluate \ourdefense against the embedding poisoning (EP) and data-free embedding poisoning (DFEP) backdoor attacks proposed recently in \cite{yang2021careful}. 
Instead of using conventional backdoor attacking methods, this kind of attacks modify a single-word embedding vector, either with data (\ie, EP method) or without data (\ie, DFEP method), to implant  backdoor. 
In EP, malicious samples are constructed following the method presented in BadNets \cite{gu2017badnets} but updating only the word embedding weight for the trigger word during backpropagation.
In DFEP, poisoning is conducted in a smaller sentence space derived from a general text corpus containing all human-written natural sentences. In our experiments, we sample sentences from the WikiText-103 corpus to create fixed-length fake samples and randomly insert the trigger word into these samples, forming a fake poisoned dataset.
We use the code shared by the authors to train backdoored BERT models for toxic input classification on Twitter data.

\begin{table}[t]
\centering
\caption{Performance of \ourdefense on toxicity detection over Twitter data with various radius $r$.}
\label{tab:Different radius of TED in NLP backdoor on toxicity detection over Twitter data}
\begin{tabular}{@{}lllllll@{}}
\toprule
\multirow{2}{*}{Attack}& \multirow{2}{*}{Metric}  & \multicolumn{4}{c}{{Radius $r$}}           & \multirow{2}{*}{AVG}     \\ \cmidrule(l){3-6} 
                                        &                         & 0.50\%  & 1.00\%  & 1.50\%  & 2.00\%  &       \\ \midrule
\multirow{3}{*}{EP} & TPR (\%)                                         & 92.50  & 92.00  & 92.00  & 92.00  & 92.12  \\
& FPR (\%)                                                             & 20.50 & 19.00 & 21.50 & 23.00 & 21.00 \\
& AUC                                                              & .9309 & .9267 & .9296 & .9280 & .9288 \\ \midrule
\multirow{3}{*}{DFEP} & TPR (\%)                                        & 95.00  & 95.00  & 94.00  & 94.00  & 94.50  \\
& FPR (\%)                                                             & 21.00 & 18.00 & 19.00 & 18.00 & 19.00 \\
& AUC                                                              & .9565 & .9528 & .9545 & .9527 & .9541 \\ \bottomrule
\end{tabular}
\end{table}

\begin{table}[t]
\centering
\caption{Performance comparison between \ourdefense, DAN, and STRIP for toxicity detection.}
\label{tab:Detection performance in NLP Backdoor on toxicity detection over Twitter data}
\begin{tabular}{@{}lllll@{}}
\toprule
Attack & Metric        & \textbf{\ourdefense}    & \textbf{DAN}     & \textbf{STRIP}   \\ \midrule
\multirow{2}{*}{EP}    & TPR (\%) & 92.12 & 93.42  & 94.99  \\ 
                       & FPR (\%) & 21.00   & 30.09 & 66.11 \\ \midrule
\multirow{2}{*}{DFEP}  & TPR (\%) & 94.50  & 93.54  & 94.99  \\
                       & FPR (\%) & 19.00   & 21.39 & 48.89 \\ \bottomrule
\end{tabular}
\end{table}



{
We implement \ourdefense with the above backdoored BERT models, and compare the detection performance of \ourdefense with other detectors, including DAN \cite{chen2022expose} and STRIP \cite{gao2019strip}.} 
Table~\ref{tab:Different radius of TED in NLP backdoor on toxicity detection over Twitter data} reports the performances of \ourdefense with various $r$, demonstrating an average AUC of 0.9288 and 0.9541  against EP and DFEP attacks, respectively. 
Moreover, as shown in Table~\ref{tab:Detection performance in NLP Backdoor on toxicity detection over Twitter data},  \ourdefense performs significantly better than DAN \cite{chen2022expose} and STRIP \cite{gao2019strip} in terms of FPR while maintaining the same level of TPR. 


%% file: sections/6_Discussions.tex
\noindent \textbf{Limitations.} Our \ourdefense is meant to detect backdoor samples for an  infected model in a blind scenario, where the attack pattern and even whether the model is infected or not are unknown. 
To this end, we approach the problem through an anomaly detection manner.
In particular, \ourdefense builds an outlier detector upon a sample set with a limited number of normal samples.  
Within this set,  a small percentage ($\alpha$) of the normal samples are identified as outliers by automatically computing an anomaly score threshold $\tau$,  and then  $\tau$ is used for detecting backdoor samples.
As a result, the subsequent detector will also likely cause some false-positive cases by treating some benign samples as outliers.
Especially when the model is not backdoored (uninfected case), it will also reject about $\alpha$ percentage of the normal inputs.
To understand such a security-utility trade-off, we conducted experiments to investigate how many percentages of normal inputs are required to be rejected under the uninfected case to keep an acceptable TPR for the infected case.
We defer the experimental details to Appendix-\ref{appendix:D}.
The results (Fig. \ref{fig:reject_alpha}) demonstrate when setting the parameter $\alpha$ ranging from 1\% to 5\%, the TPRs in the infected cases under all three datasets are all above 92.5\%.
Furthermore, other than setting the hard parameter $\alpha$, we adopt the Z-score-based method, a more advanced approach determining the reject threshold that is robust regardless of whether the model is infected or not on CIFAR-10.
The  threshold $\tau$ is computed by setting to reject not strictly $\alpha$ percentage of outlier, but the samples that exceed four standard deviations ($4 \times \sigma$) from the mean of the feature distribution under Gaussian modeling. 
In this way, only the samples that deviate from the distribution to a  certain extent will be rejected.
This method results in a TPR of 92.4\% on the infected model.  
Notably, the uninfected model exhibits an FPR of only 0.7\%,  which is fairly acceptable.

Fundamentally, \ourdefense leverages the topological evolution dynamic to build the detector, whose working principle relies on the assumption that backdoor samples should have a drastically different trajectory relative to the benign samples of target label $t$.
In our evaluation, this assumption seems to hold in general, even under some very sophisticated attacks. 
In particular, with our proposed \ourattack attack that combines the best elements of existing attack approaches with hard-to-detect properties, the backdoor samples still show expected irregularities in their trajectory.
Furthermore, despite designing multiple adaptive attacks to challenge this assumption, none have succeeded. 
However, we acknowledge that future research may reveal novel attacks that invalidate this assumption if they successfully render the activations of backdoor samples to be inside the open balls of benign samples of target label $t$ as they propagates in the network and still manage to inject the backdoor.

\noindent \textbf{Future work.} In this work, we propose to utilize the evolution dynamic for backdoor sample detection. 
Specifically, \ourdefense captures the evolution dynamics by analyzing the rank sequence of the nearest samples belonging to the predicted class. 
This approach has demonstrated remarkable performance in countering existing attack methods thus far.
However, we believe that further advancements can be made by improving the modeling of the evolution dynamics.
Merely relying on the rank of the nearest sample from the predicted class as a measure of ``closeness" is a straightforward approach but may not be optimal. Therefore, future research could explore alternative methods to assess the proximity of an input sample to the distribution of samples from the target class at each layer. 
By examining the ``closeness" in this manner, we anticipate that the performance of our approach could be further enhanced.

%% file: sections/7_Conclusion.tex
Backdoor attacks present a critical security threat to the DNN model supply chain, and many defensive mechanisms have been proposed to address this threat. Our work identified that existing backdoor detectors' success relies on the assumption of trigger strategy such that the latent representations of benign and malicious samples can be separated in the metric space. Experimental results confirmed that SOTA backdoor detectors built on this assumption are ineffective against our proposed \ourattack, a sophisticated backdoor attack that blends both the source-specific and the dynamic-trigger strategies. 
To overcome this limitation, we turned our attention to the analysis in the topological space and proposed \ourdefense, capitalizing on the inherent differences in the evolutionary dynamics of topological structures between benign and malicious samples.
The effectiveness of \ourdefense is corroborated through extensive experimental results on CV and NLP tasks. 
By adopting a novel perspective that considers the topological structure, this work represents a significant step forward in understanding and mitigating backdoors in deep learning. 


%% file: sections/8_appendix.tex

\appendices \label{Appendix}

\section{Detailed Settings and More Experimental Results}
\label{appendix:A}

\begin{table}[b!]
    \centering
    \caption{Network Architecture for \ourattack on MNIST}
    \label{tab:mnist_model_SSDT}
    \begin{tabular}{lcccc}
        \toprule
        Layer & Output Size & Kernel Size & Stride & Activation \\
        \midrule
        Conv2d-1       & $32 \times 24 \times 24$ & $5 \times 5$ & 1 & ReLU \\
        Dropout-3      & $32 \times 24 \times 24$ & - & - & - \\
        MaxPool2d-4    & $32 \times 12 \times 12$ & $2 \times 2$ & 2 & - \\
        Conv2d-5       & $64 \times 8 \times 8$   & $5 \times 5$ & 1 & ReLU \\
        Dropout-7      & $64 \times 8 \times 8$   & - & - & - \\
        MaxPool2d-5    & $64 \times 4 \times 4$   & $2 \times 2$ & 2 & - \\
        Linear-6       & 512                      & - & - & ReLU \\
        Dropout-8      & 512                      & - & - & - \\
        Linear-9       & 10                       & - & - & - \\
        \bottomrule
    \end{tabular}

\end{table}

\noindent\textbf{Settings:}
For the implementation of \ourattack, we utilize three datasets MNIST, CIFAR-10, and GTSRB. For MNIST, we employ a 2-layer CNN as shown in Table \ref{tab:mnist_model_SSDT}.
For CIFAR-10 and GTSRB, we use Pre-activation ResNet18 \cite{he2016identity}. 
In Algorithm~\ref{algo:ssid}, the sampling rates $\rho, \rho_b, \rho_l, \rho_{ct}$ associated with no-trigger samples, victim-triggered samples, non-victim but triggered samples, and cross-triggered samples are respectively set as 
 $\rho = \frac{c}{
 c+2}$,  
$\rho_b=\frac{1}{ c+2}$ and $\rho_l+\rho_{ct}=\frac{1}{ c+2}$, where $c$ is the number of classes.

For the implementation of \ourdefense, we randomly sample 20 clean images for each class from the test sets of MNIST and CIFAR-10 (both with 10 different classes), respectively, and we randomly sample 1,000 clean images in total from the test set of GTSRB (43 different classes).
The {contamination rate (\ie, FPR)} parameter $\alpha$ in  Algorithm~\ref{algo:ted} is set to 5\% unless otherwise specified.

\begin{figure*}[t!]
    \centering

     \begin{subfigure}[b]{0.4\textwidth}
        \centering
        \includegraphics[width=\textwidth]{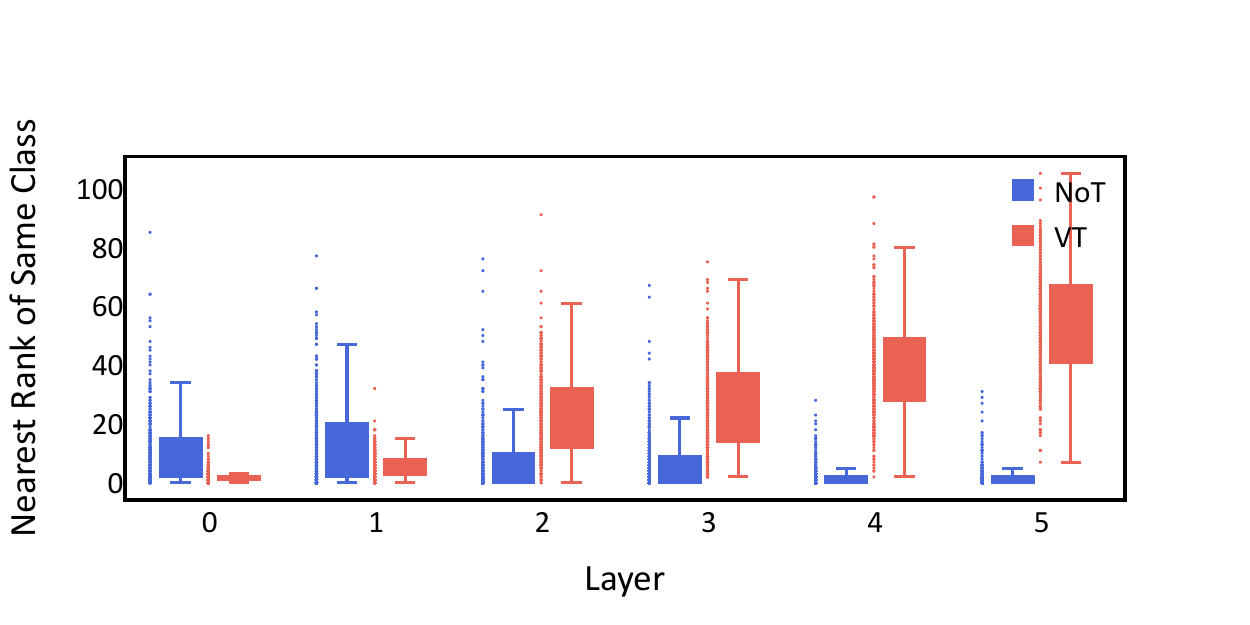}
        \caption{CIFAR-10}
        \label{fig:TopologyPersistenceDiagram_CIFAR10_TaCT}
    \end{subfigure}
    \begin{subfigure}[b]{0.4\textwidth}
        \centering
        \includegraphics[width=\textwidth]{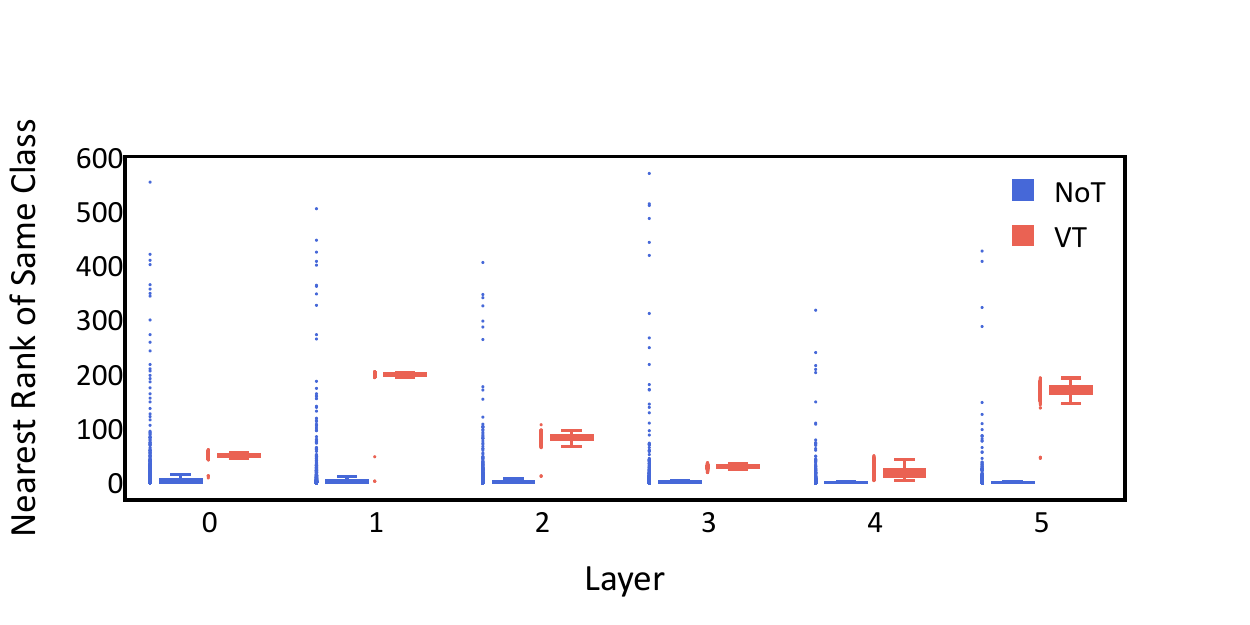}
        \caption{GTSRB}
  \label{fig:TopologyPersistenceDiagram_GTSRB_TaCT}
    \end{subfigure}
    \caption{{Box plot of topological feature vectors 
 under TaCT~\cite{tang2021demon}}.}
    \label{fig:TopologyPersistenceDiagram_TaCT}
\end{figure*}

\begin{figure*}[t!]
    \centering
        \begin{subfigure}[b]{0.7\textwidth}
        \centering
        \includegraphics[width=\textwidth]{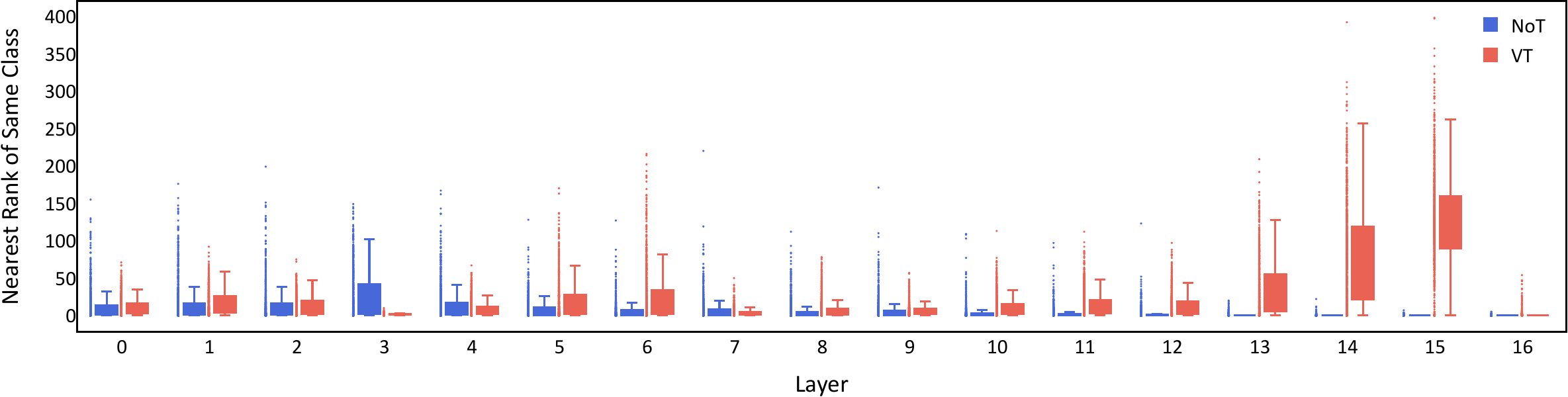}
        \caption{CIFAR-10}\label{fig:TopologyPersistenceDiagram_cifar10_DynamicInput}
    \end{subfigure}
    \begin{subfigure}[b]{0.7\textwidth}
        \centering
        \includegraphics[width=\textwidth]{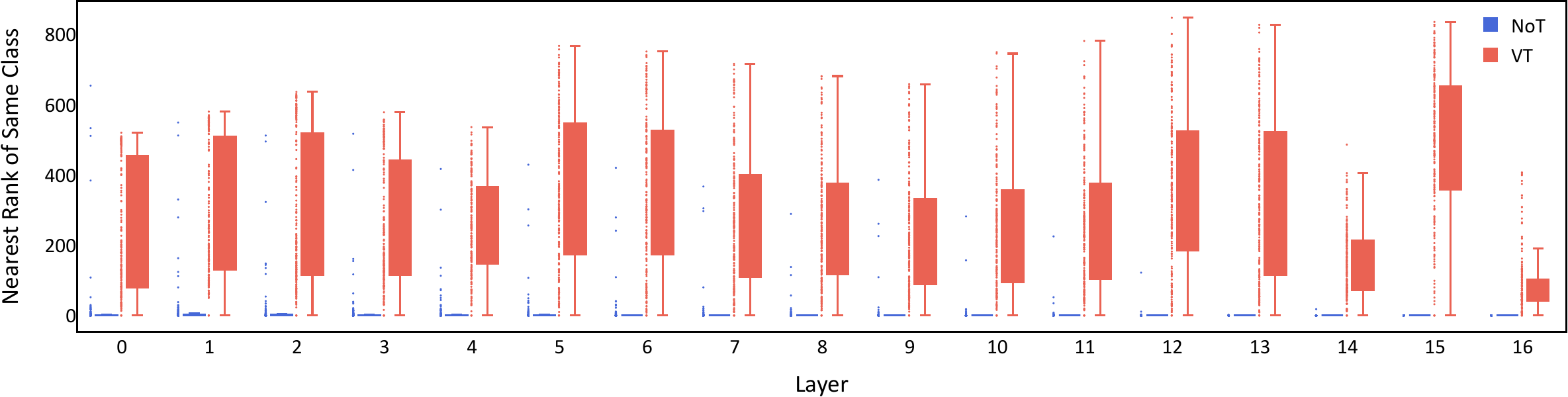}
        \caption{GTSRB}\label{fig:TopologyPersistenceDiagram_GTSRB_DynamicInput}
    \end{subfigure}       

    \caption{Box plot of topological feature vectors 
 under the dynamic trigger attack \cite{nguyen2020input}.}
    \label{fig:TopologyPersistenceDiagram_DynamicInput}
\end{figure*}

\begin{figure*}[t!]
    \centering
        \begin{subfigure}[b]{0.7\textwidth}
        \centering
        \includegraphics[width=\textwidth]{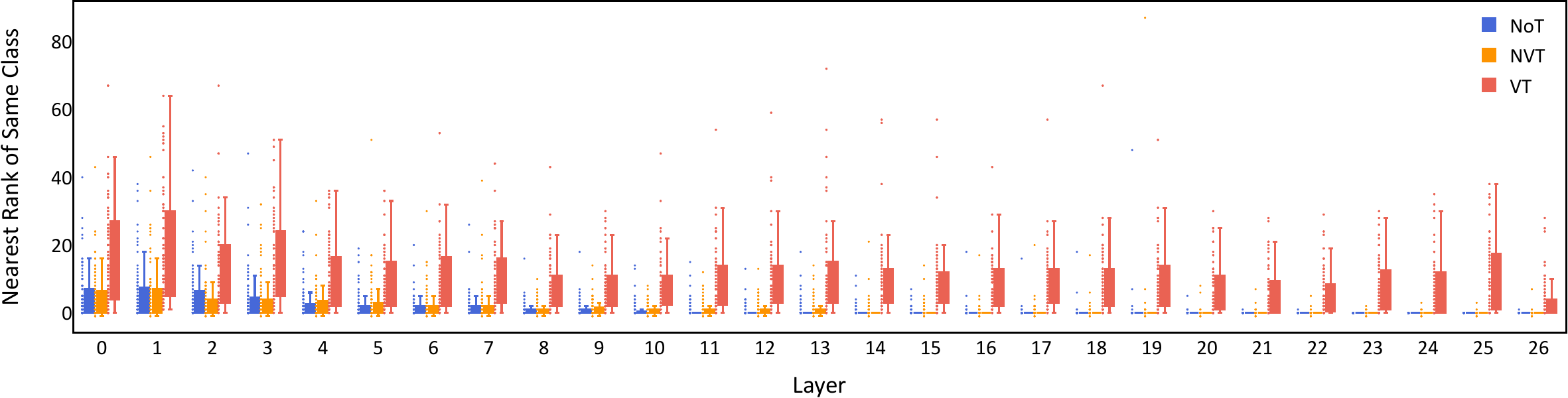}
        \caption{PubFig}\label{fig:pubfig_TED}
    \end{subfigure}    
     \hfill
     \begin{subfigure}[b]{0.7\textwidth}
        \centering
        \includegraphics[width=\textwidth]{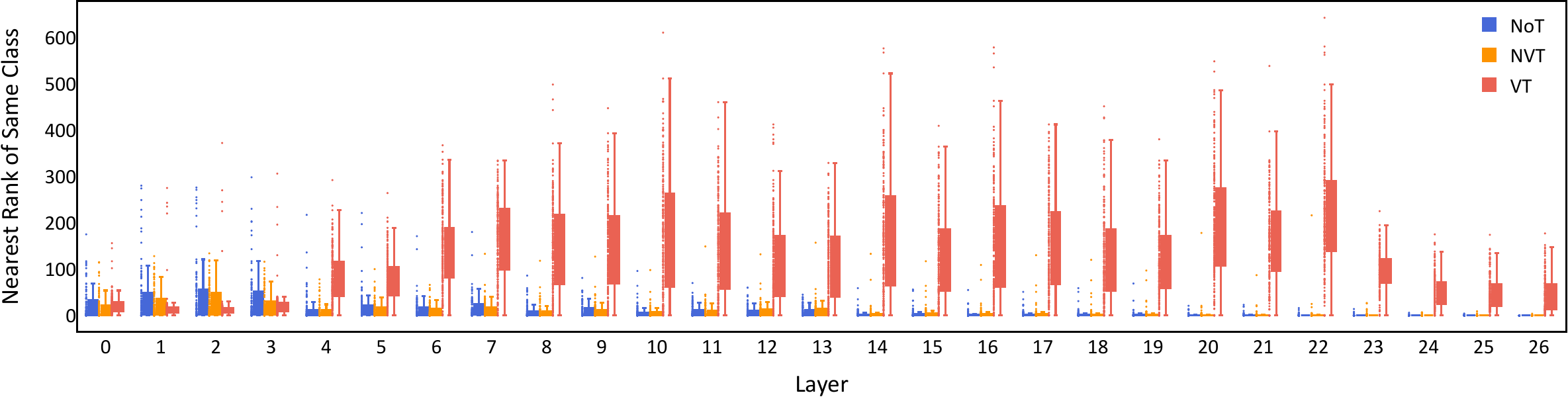}
        \caption{ImageNet}\label{fig:imagenet_TED}
    \end{subfigure}

    \caption{Box plot of topological feature vectors on Pubfig and ImageNet under \ourattack. }
    \label{fig:TopologyPersistenceDiagram_pubfig_imagenet}
\end{figure*}

\noindent\textbf{More results:} 
Fig.~\ref{fig:TopologyPersistenceDiagram_DynamicInput}  box-plots the topological feature vectors under a dynamic-trigger backdoor attack proposed in \cite{nguyen2020input}. Similarly, Fig.~\ref{fig:TopologyPersistenceDiagram_TaCT} box-plots the topological feature vectors under the source-specific backdoor attack TaCT proposed in \cite{tang2021demon}.
From both figures, it is easy to see that the feature vectors between NoT samples and VT samples are clearly separable.
Under complex dataset scenarios, as depicted in Fig. \ref{fig:TopologyPersistenceDiagram_pubfig_imagenet}, VT samples exhibit more variability on their way towards the predicted class, whereas the NoT and NVT samples consistently approach the predicted class from the front to the back in both the PubFig and ImageNet datasets.

\section{Details of Adaptive Attacks}
\label{appendix:SUPPadatpativeloss}

Here, we present the details of the three loss functions for adaptive attacks. These functions serve the purpose of mimicking the topological features of benign samples for adaptive attacks of \ourdefense. 

As mentioned above, $\mathcal{L}_1$ employs the large margin nearest neighbor classification  method \cite{weinberger2009distance} to minimize the Euclidean distance between poisoned activations within the same class while maximizing it between different classes, hence mimicking the topological feature of a benign sample. In particular, $\mathcal{L}_1$ is defined as 
\begin{IEEEeqnarray}{rCl} 
\mathcal{L}_1 &=& \frac{1}{\#D_b}\sum\nolimits_{i=1}^{\#D_b} \max\left(0, \min_i(\{d_{T,i}\}) - \min_i(\{d_{O,i}\})\right), 
\nonumber
\end{IEEEeqnarray}
where $\#D_b$ is the number of poisoned samples, $d_{T,i}$ and $d_{O,i}$ represent the pairwise Euclidean distances between the $i$-th poisoned activation and other activations of the target class and other classes, respectively. 

$\mathcal{L}_2$ employs the idea of the K-means algorithm \cite{MacQueen1967SomeMF} to reduce the distance from a given point (the poisoned activation) to the geometric centroid of the target class, which is defined as 
\begin{IEEEeqnarray}{rCl} 
\mathcal{L}_2 = \frac{1}{\#D_b}\sum\nolimits_{i=1}^{\#D_b} ||f_{N-1}\circ \cdots \circ f_1(x_i) - c_t||_2, \nonumber
\end{IEEEeqnarray}
where $c_t$ represents the centroid of the target class and $f_{N-1}\circ \cdots \circ f_1(x_i)$ represents the $i$-th poisoned activation.

$\mathcal{L}_3$ employs the shallow-layer weight poisoning method in \cite{li2021backdoor} to make topological features associated with deeper layers appear normal. In particular, $\mathcal{L}_3$ is defined as 
\begin{IEEEeqnarray}{rCl} 
\mathcal{L}_3 = && \frac{1}{\#D_b*\#\bar{D}}
\sum\nolimits_{i=1}^{\#D_b} \sum\nolimits_{j=1}^{\#\bar{D}} ||f_{k}\circ \cdots \circ f_1(x_i) \nonumber \\ &&- f_{k}\circ \cdots \circ f_1(x^t_j)||_2, \nonumber
\end{IEEEeqnarray}
where $\bar{D}$ is a small number of samples from the target class, and $f_{k}\circ \cdots \circ f_1$ are the first $k$ layers of $f$.

\section{Detecting Backdoor in Shallow  Networks Without Convolution}
\label{appendix:SSDTinShallowmodels}
We conduct experiments using a backdoored ShallowNet on MNIST. The ShallowNet consists of two fully connected layers. The first layer takes input features of size 1,024 and maps them to an intermediate representation with 128 units. The second layer takes this 128-unit representation and maps it to the final output, which consists of 10 classes, corresponding to the 10 digits in MNIST. The {TaCT backdoor method~\cite{tang2021demon} is used, and the backdoored ShallowNet achieves 97.6\% accuracy on NoT samples and 100\% on VT samples.} As discussed in Sec.~\ref{subsec:AblationStudy}, we take the two Linear layers to extract the topological features for \ourdefense. As listed in Table~\ref{table:Defending Performance on Shallow NN in MNIST}, we observe that even with only two Linear layers, \ourdefense is still able to perform well. The precision of \ourdefense is 98.76\% and the accuracy is 99.13\%.



\begin{table}[t]
\centering
\caption{Detection performance of ShallowNet on MNIST (TPR for VT: 99.50\%, FPR for NVT: 2.50\%, FPR for NoT: 0.00\%).}
\label{table:Defending Performance on Shallow NN in MNIST}
\begin{tabular}{@{}cc|cc@{}}
\toprule
Metric & Value &Metric & Value        \\ \midrule
Precision         & 98.76\%     &Accuracy  & 99.13\%       \\ \bottomrule
\end{tabular}
\end{table}

\section {Detection threshold $\tau$ selection.}
\label{appendix:D}

\begin{figure}[h!]
    \includegraphics[width=0.45\textwidth]{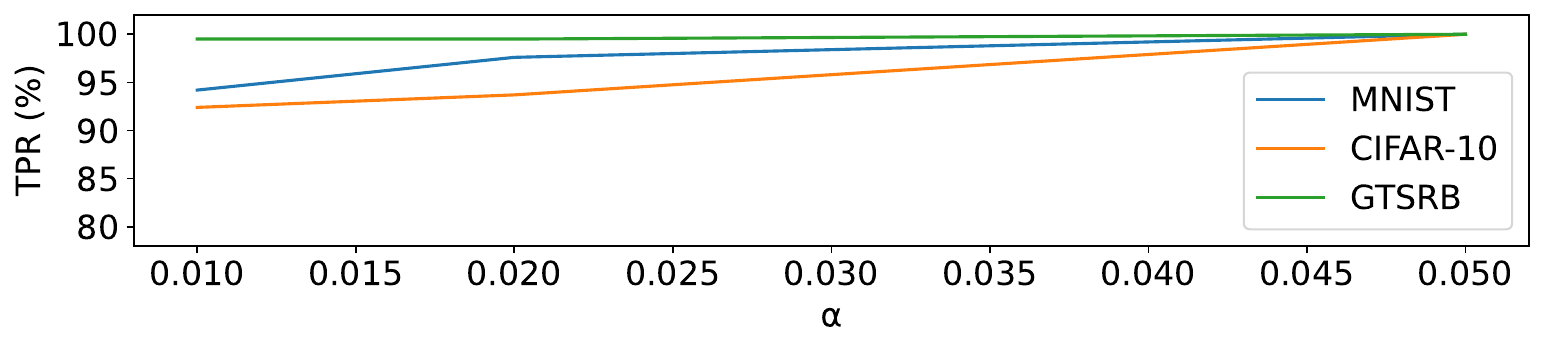}
    \caption{Ablation on 
{the parameter $\alpha$}, with plots of TPR of each $\alpha$ on three datasets.}
    \label{fig:reject_alpha}
\end{figure}

A Z-score-based outlier detection rule is applied to both a benign model and a \ourattack-backdoored model on CIFAR-10. Inputs are classified as ``positive detection” if the \ourdefense anomaly score exceeds four standard deviations (4×$\sigma$) from the feature distribution mean, setting the threshold $\tau$. This method results in a TPR of 92.4\% on the backdoored model, indicating the successful detection of most poisoned inputs with minimal false positives. Notably, the benign model exhibits an FPR of only 0.7\%, an acceptable rate demonstrating the method's precision.

Also, an ablation study on the reject parameter {$\alpha$}, ranging from 1\% to 5\%, reveals that all TPRs across the three datasets exceed 92.5\%, as illustrated in Fig.~\ref{fig:reject_alpha}.

